\documentclass[english,british]{paper}
\usepackage[T1]{fontenc}
\usepackage[latin9]{inputenc}
\usepackage{geometry}
\usepackage{float}
\usepackage{bm}
\usepackage{amsthm}
\usepackage{amsmath}
\usepackage{amssymb}
\usepackage{graphicx}
\usepackage{esint}
\usepackage[authoryear]{natbib}

\makeatletter

\floatstyle{ruled}
\newfloat{algorithm}{tbp}{loa}
\providecommand{\algorithmname}{Algorithm}
\floatname{algorithm}{\protect\algorithmname}

\numberwithin{equation}{section}
\numberwithin{figure}{section}

\usepackage{dsfont}

\@ifundefined{showcaptionsetup}{}{%
 \PassOptionsToPackage{caption=false}{subfig}}
\usepackage{subfig}
\makeatother

\usepackage{babel}
\addto\captionsenglish{\renewcommand{\algorithmname}{Algorithm}}

\begin{document}

\title{Expectation-Propagation for Likelihood-Free Inference}

\author{Simon Barthelmé%
\thanks{Université de Genève. Faculté de Psychologie et de Sciences de l'Education,
Unimail CH-1211 Genève 4, Switzerland. simon.barthelme@unige.ch%
} \\
Nicolas Chopin%
\thanks{CREST/ENSAE. 3, Avenue Pierre Larousse 92245 Malakoff CEDEX France.
nicolas.chopin@ensae.fr%
} }
\maketitle
\begin{abstract}
Many models of interest in the natural and social sciences have no
closed-form likelihood function, which means that they cannot be treated
using the usual techniques of statistical inference. In the case where
such models can be efficiently simulated, Bayesian inference is still
possible thanks to the Approximate Bayesian Computation (ABC) algorithm.
Although many refinements have been suggested, ABC inference is still
far from routine. ABC is often excruciatingly slow due to very low
acceptance rates. In addition, ABC requires introducing a vector of
``summary statistics'' $\bm{s}(\bm{y})$, the choice of which is
relatively arbitrary, and often require some trial and error, making
the whole process quite laborious for the user. 

We introduce in this work the EP-ABC algorithm, which is an adaptation
to the likelihood-free context of the variational approximation algorithm
known as Expectation Propagation \citep{minka2001expectation}. The
main advantage of EP-ABC is that it is faster by a few orders of magnitude
than standard algorithms, while producing an overall approximation
error which is typically negligible. A second advantage of EP-ABC
is that it replaces the usual global ABC constraint $\left\Vert \bm{s}(\bm{y})-\bm{s}(\bm{y}^{\star})\right\Vert \leq\varepsilon$,
where $\bm{s}(\bm{y}^{\star})$ is the vector of summary statistics
computed on the whole dataset, by $n$ local constraints of the form
$\left\Vert s_{i}(y_{i})-s_{i}(y_{i}^{\star})\right\Vert \leq\varepsilon$
that apply separately to each data-point. In particular, it is often
possible to take $s_{i}(y_{i})=y_{i}$, making it possible to do away
with summary statistics entirely. In that case, EP-ABC makes it possible
to approximate directly the evidence (marginal likelihood) of the
model. 

Comparisons are performed in three real-world applications which are
typical of likelihood-free inference, including one application in
neuroscience which is novel, and possibly too challenging for standard
ABC techniques. 

Key-words: Approximate Bayesian Computation; Composite Likelihood;
Expectation Propagation; Likelihood-Free Inference; Quasi-Monte Carlo. 
\end{abstract}
\global\long\def\bt{\bm{\theta}}
\global\long\def\bmu{\bm{\mu}}
\global\long\def\bl{\bm{\lambda}}
\global\long\def\be{\bm{\eta}}
\global\long\def\E{\mathbb{E}}
\global\long\def\by{\bm{y}}
\global\long\def\bys{\bm{y^{\star}}}
\global\long\def\bS{\bm{\Sigma}}
\global\long\def\N{\mathcal{N}}
\global\long\def\I{\mathds{1}}

\section{Introduction}

In natural and social sciences, one finds many examples of probabilistic
models whose likelihood function is intractable. This includes most
models of noisy biological neural networks \citep{GerstnerKistler:SpikingNeuronModels},
some time series and choice models in Economics \citep{Train:DiscreteChoiceSimulation},
phylogenetic models in evolutionary Biology \citep{Beaumont:ABCinEvolutionEcology},
spatial extremes in Environmental Statistics \citep{SpatialExtremesReview},
among others. That the likelihood is intractable is unfortunate, because
one would still like to perform the usual statistical tasks of parameter
inference and model comparison, and the traditional statistical tool-kit
assumes that the likelihood function is either directly available
or can be made so by introducing latent variables. This explains that
researchers have often had to content themselves with semi-quantitative
analyses showing that a model could reproduce some aspect of an empirical
phenomenon for some values of the parameters; for two representative
examples from vision science, see \citet{Nuthmann:CRISP}, \citet{Brascamp:TimeCourseBinocularRivalry}. 

A breakthrough was provided by the work of \citet{tavare1997inferring},
\citet{pritchard1999ABC} and \citet{beaumont2002ABC}, in the form
of the Approximate Bayesian Computation (ABC) algorithm, which enables
Bayesian inference in the likelihood-free context. (See also \citet{diggle1984monte}
and \citet{rubin1984bayesianly} for early versions of ABC.) Assuming
some model $p(\bm{y}^{*}|\bm{\theta})$ for the data $\by^{\star}$,
and a prior $p(\bm{\theta})$ over the parameter $\bt\in\Theta$,
the ABC algorithm iterates the following steps: 
\begin{enumerate}
\item Draw $\bm{\theta}$ from the prior, $\bt\sim p(\bt)$.
\item Draw a dataset $\bm{y}$ from the model conditional on $\bm{\theta}$,
$\by|\bt\sim p(\by|\bt)$.
\item If $d(\bm{y},\bm{y^{*}})\leq\epsilon$, then keep $\bm{\theta}$,
otherwise reject. 
\end{enumerate}
and therefore produces samples from the so-called ABC posterior:

\begin{equation}
p_{\epsilon}\left(\bt|\bm{y}^{*}\right)\propto p(\bm{\theta})\int p(\bm{y}|\bm{\theta})\I_{\left\{ d(\bm{y},\bm{y^{\star})\leq\epsilon}\right\} }\, d\bm{y}.\label{eq:ABCpost}
\end{equation}

The pseudo-distance is usually taken to be $d(\by,\by^{\star})=||\bm{s}\left(\bm{y}\right)-\bm{s}\left(\bm{y}^{*}\right)||$,
for some norm $||\cdot||$, where $\bm{s}(\bm{y})$ is a vector of
summary statistics, for example some empirical quantiles or moments
of \textbf{$\bm{y}$}. Unless $\bm{s}$ is sufficient, the approximation
error does\textbf{ }not vanish as $\epsilon\rightarrow0$, as $p(\bt|\bm{s}(\bm{y^{\star}}))\neq p(\bt|\bm{y^{\star}})$.
In that respect, the ABC posterior suffers from two levels of approximation:
a nonparametric error, akin to the error of kernel density estimation,
and where $\epsilon$ plays the role of a bandwidth \citep[see e.g. ][]{Blum:ABCNonParametricPerspective},
and a bias introduced by the summary statistics $\bm{s}$. The more
we include in $\bm{s}\left(\bm{y}\right)$, the smaller the bias induced
by $\bm{s}$ should be. On the other hand, as the dimensionality of
$\bm{s}(\bm{y})$ increases, the lower the acceptance rate will be.
We would then have to increase $\epsilon$, which leads to an approximation
of lower quality.

Thus, ABC requires in practice some more or less arbitrary compromise
between what summary statistics to include and how to set $\epsilon.$
To establish that the results of the inference are somewhat robust
to these choices, many runs of the algorithm may be required. Although
several variants of the original ABC algorithm that aim at increasing
acceptance rates exist (e.g. \citealp{beaumont2002ABC}), the current
state of the matter is that an ABC analysis is very far from routine
use because it may take days to tune on real problems. The semi-automatic
method for constructing summary statistics recently proposed by \citet{fearnheadprangle}
seems to alleviate partly these problems, but it still requires at
least one, and sometimes several pilot runs.

In this article we introduce EP-ABC, an adaptation of the Expectation
Propagation (EP) algorithm \citep[Chap. 10]{minka2001expectation,Bishop:book}
to the likelihood-free setting. The main advantage of EP-ABC is that
it is much faster than previous ABC algorithms: typically, it provides
accurate results in a few minutes, whereas a standard ABC algorithm
may need several hours, or even days. EP-ABC requires that the data
$\bm{y}^{\star}$ may be decomposed into $n$ ``chunks'', $y_{1}^{\star},\ldots,y_{n}^{\star}$
(of possibly different dimensionality or support), in such a way that
is possible to simulate sequentially the $n$ chunks; i.e. one is
able to simulate from $p(y_{i}^{\star}|y_{1:i-1}^{\star},\bt)$ for
each $i$. More precisely, EP-ABC builds an EP approximation of the
following type of ABC posterior distributions: 
\begin{equation}
p_{\epsilon}(\bt|\by^{\star})\propto p(\bt)\prod_{i=1}^{n}\left\{ \int p(y_{i}|y_{1:i-1}^{\star},\bt)\I_{\left\{ \Vert s_{i}(y_{i})-s_{i}(y_{i}^{\star})\Vert\leq\varepsilon\right\} }\, dy_{i}\right\} \label{eq:kernel_function}
\end{equation}
where $s_{i}(y_{i})$ is a summary statistic specific to chunk $i$,
and with the convention that $p(y_{1}|y_{1:0}^{\star},\bt)=p(y_{1}|\bt)$.
Given the way it operates, we shall see that EP-ABC essentially replaces
the initial, possibly difficult ABC problem, by $n$ ABC simpler (i.e.
lower-dimensional) ABC problems. In particular, it may be easier to
construct a set of ``local'' summary statistics $s_{i}(y_{i})$
for each chunk $y_{i}$ rather than a global set $\bm{s}(\bm{y})$
the whole dataset $\by$, in such a way that the probability that
$\Vert s_{i}(y_{i})-s_{i}(y_{i}^{\star})\Vert\leq\varepsilon$ is
not too small. 

Of course, not all ABC models are amenable to the factorisation in
\eqref{eq:kernel_function}, at least directly. We shall discuss this
point more in detail in the paper. At the very least, factorisable
models include situations with $n$ repeated experiments, which are
not easily tackled through standard ABC, because of the difficulty
to define some vector $\bm{s}(\by)$ which summarises these $n$ experiments.
\citet{bazin2010likelihood} discusses this point, in the context
of genetic data collected over a large number of loci, and recommend
instead to construct loci-specific summary statistics, and combine
results from loci-specific ABC exercises.

In certain problems such that $y_{i}$ is of low dimension, it may
be even possible to take $s_{i}(y_{i})=y_{i}$. In that case, EP-ABC
provides an EP approximation of a summary-less ABC posterior; that
is, when $\epsilon\rightarrow0$ (under appropriate measure-theoretic
conditions), $ $$p_{\epsilon}(\bt|\by^{\star})\rightarrow p(\bt|\by^{\star})$,
the true posterior. In such situations, EP-ABC also provides an approximation
of the evidence $p(\by^{\star})$, which is a very useful quantity
for model comparison. \citet{dean2011parameter} show that, in the
specific context of hidden Markov models such that the observation
density is intractable (as this work was motivated by the filtering
ABC algorithm developed in \citet{filteringabc}), the ABC error of
a summary-less ABC algorithm should not be interpreted as a non-parametric
error, but rather as a misspecification error, where the misspecified
model is the true model, but with observations corrupted with noise.
This misspecification implies a bias of order $\varepsilon$. Thus,
in addition to being more convenient for the user, as it avoids specifying
some summary statistics, summary-less ABC does not seem to suffer
from the curse of dimensionality (in the dimension of $\bm{s}(\bm{y})$)
of standard ABC. Allowing summary-less ABC inference in certain cases
seems to be another advantage of EP-ABC. 

We start with a generic description of the EP algorithm, in Section
\ref{sec:Principle_of_EP}, and explain in Section \ref{sec:ABC-EP}
how it can be adapted to the likelihood-free setting. We explain in
Section \ref{sec:IID_case} how EP-ABC can be made particularly efficient
when data-points are IID (independent and identically distributed).
Section \ref{sec:Numerical_examples} contains three case studies
drawn from finance, population ecology, and vision science. The two
first examples are borrowed from already known applications of ABC,
and illustrate to which extent EP-ABC may outperform standard ABC
techniques in realistic scenarios. To the best of our knowledge, our
third example from vision science is a novel application of likelihood-free
inference, which, as which shall argue, seems too challenging for
standard ABC techniques. 

Section \ref{sec:Difficult-posteriors} discusses how EP-ABC may cope
with difficult posteriors through two additional examples. Section
\ref{sec:Extensions} discusses possible extensions of EP-ABC; in
particular, for non-factorisable likelihoods. In such cases, one may
replace the likelihood by some factorisable approximation, such as
a composite likelihood. Section \ref{sec:Conclusion} concludes. 

We use the following notations throughout the paper: bold letters
refer to vectors or matrices, e.g. $\bt$, $\bl$, $\bm{\Sigma}$
and so on. We also use bold face to distinguish complete sets of observations,
i.e. $\by$ or $\by^{\star}$, from their components, $y_{i}$, and
$y_{i}^{\star}$, $i=1,\ldots n$, although we do not assume that
these components are necessarily scalar. For sub-vectors of observations,
we use the colon notation: $y_{1:i}=(y_{1},\ldots,y_{i})$. The notation
$\Vert\cdot\Vert$ refers to a generic norm, and $\Vert\cdot\Vert_{2}$
refers to the Euclidean norm. The Kullback-Leibler divergence between
probability densities $\pi$ and $q$ is denoted as 
\[
KL(\pi\Vert q)=\int\pi(\bt)\log\left(\frac{\pi(\bt)}{q(\bt)}\right)\, d\bt.
\]
The letter $p$ always refers to probability densities concerning
the model; i.e. $p(\bt)$ is the prior, $p(y_{1}|\bt)$ is the likelihood
of the first observation, and so on. Transpose of a matrix $\bm{A}$
is denoted $\bm{A}^{t}$, and the diagonal matrix with diagonal elements
given by vector $\bm{V}$ is $\mathrm{diag}(\bm{V})$.

\section{Expectation Propagation\label{sec:Principle_of_EP}}

Expectation Propagation \citep[EP,][]{minka2001expectation} is an
algorithm for variational inference, a class of techniques that aim
at finding a tractable probability distribution $q(\bt)$ that best
approximates an intractable target density $\pi(\bt)$. One way to
formulate this goal is as finding the member of some parametric family
$\mathcal{Q}$ that is in some sense closest to $\pi\left(\bt\right)$,
where ``closest'' is defined by some divergence between probability
distributions. A distinctive feature of EP is that it based on the
divergence $KL(\pi||q)$, whereas many variational methods \citep[e.g. Variational Bayes, see Chap. 10 of ][]{Bishop:book}
try to minimize $KL(q||\pi)$. For a good discussion of the relative
merits of each divergence, see \citet[p. 468-470]{Bishop:book} and
\citet{Minka:DivMeasuresMP}. As a brief summary, minimizing $KL(q||\pi)$
tends to produce approximation that are too compact; see also \citet{wang2005inadequacy}.
The divergence $KL(\pi||q)$ does not suffer from this drawback, and
is perhaps more appealing from a statistical point of view (if only
because maximum likelihood estimation is based on the corresponding
contrast), but on the other hand, minimizing $KL(\pi||q)$ when $\pi$
is multimodal tends to produce an approximation that covers all the
modes, and therefore has too large a support. We shall return to this
point.

\subsection{Assumptions of Expectation Propagation\label{sub:Assumptions-of-EP}}

EP assumes that the target density $\pi(\bt)$ decomposes into a product
of simpler factors 
\begin{equation}
\pi(\bt)=\frac{1}{Z_{\pi}}\prod_{i=0}^{n}l_{i}\left(\bt\right)\label{eq:EPdecomposition}
\end{equation}
and exploits this factorisation in order to construct a sequence of
simpler problems. For instance, $l_{0}(\bm{\theta})$ may be the prior
$p(\bt)$, $l_{i}(\bt)=p(y_{i}|y_{1:i-1},\bt)$ if $\by$ is a set
of $n$ observations $y_{i}$, with the convention that $p(y_{1}|y_{1:0},\bt)=p(y_{1}|\bt)$,
and $Z_{\pi}=p(\by).$

EP uses an approximating distribution with a similar structure:

\begin{equation}
q(\bt)\propto\prod_{i=0}^{n}f_{i}(\bt)\label{eq:q-factorised}
\end{equation}
where the $f_{i}$'s are known as the ``sites''. In a spirit close
to a coordinate-descent optimization algorithm, each site is updated
in turn, while the $n$ other sites are kept fixed. 

For the sake of conciseness, we focus in this paper on Gaussian sites,
expressed under their natural parametrisation: $f_{i}\left(\bt\right)=\exp\left(-\frac{1}{2}\bt^{t}\bm{Q_{i}}\bt+\bm{r}_{i}^{t}\bt\right)$,
where $\bm{Q}_{i}$ and $\bm{r}_{i}$ are called from now on the \emph{site
parameters. }This generates the following Gaussian approximation $q$:

\begin{equation}
q(\bt)\propto\exp\left\{ -\frac{1}{2}\bt^{t}\left(\sum_{i=0}^{n}\bm{Q}_{i}\right)\bt+\left(\sum_{i=0}^{n}\bm{r}_{i}\right)^{t}\bt\right\} .\label{eq:q-Gaussian-natural}
\end{equation}
In addition, we assume that the true prior $p(\bt)$ is Gaussian,
with natural parameters $\bm{Q}_{0}$ and $\bm{r}_{0}$. In that case,
the site $f_{0}$ is kept equal to the prior, and only the sites $f_{1}$
to $f_{n}$ are updated. 

We note however that EP may easily accommodate a non-Gaussian prior
(by simply treating the prior as additional factor to be approximated),
or other types of parametric sites. In fact, the site update described
in the following section may be easily adapted to any exponential
family; see \citet{Seeger:EPExpFam}.

\subsection{Site update\label{sub:A-geometrical-view}}

Suppose that \eqref{eq:q-factorised} is the current approximation,
and one wishes to update site $i$. This is done by creating a ``hybrid''
distribution, obtained by substituting site $i$ with the true likelihood
contribution $l_{i}(\bt)$:

\begin{equation}
h\left(\bt\right)\propto q_{-i}\left(\bt\right)l_{i}(\bt),\quad q_{-i}(\bt)=\prod_{j\neq i}f_{j}(\bt).\label{eq:hybrid}
\end{equation}
For Gaussian sites, this leads to: 

\[
h(\bt)\propto l_{i}\left(\bt\right)\exp\left(-\frac{1}{2}\bt^{t}\bm{Q}_{-i}\bt+\bm{r}_{-i}^{t}\bt\right),\quad\bm{Q}_{-i}=\sum_{j\neq i}\bm{Q}_{j},\quad\bm{r}_{-i}=\sum_{j\neq i}\bm{r}_{j}.
\]

The new value of site $f_{i}$ is then obtained by minimising with
respect to $f_{i}$ the Kullback-Leibler pseudo-distance $KL(h_{i}||q)$
between the hybrid and the Gaussian approximation $q$ (again keeping
the $f_{j}$'s fixed). When Gaussian sites are used, this minimisation
is equivalent to taking $q$ to be the Gaussian density with moment
parameters that match those of the hybrid distribution 
\begin{eqnarray}
Z_{h} & = & \int q_{-i}\left(\bt\right)l_{i}(\bt)\, d\bt,\nonumber \\
\bm{\mu}_{h} & = & \frac{1}{Z_{h}}\int\bt q_{-i}\left(\bt\right)l_{i}(\bt)\, d\bt,\nonumber \\
\bm{\Sigma}_{h} & = & \frac{1}{Z_{h}}\int\bt\bt^{t}q_{-i}\left(\bt\right)l_{i}(\bt)\, d\bt-\bm{\mu}_{h}\bm{\mu}_{h}^{t}.\label{eq:EP_moments_Gaussiancase}
\end{eqnarray}

A key observation at this stage is that the feasibility of EP is essentially
dictated by how easily the moments above may be computed. These moments
may be interpreted as the moments of a posterior distribution, based
on a Gaussian prior, with natural parameters $\bm{Q}_{-i}$ and $\bm{r}_{-i}$,
and a likelihood consisting of a single factor $l_{i}(\bt)$. 

Finally, from these moment parameters, one may recover the natural
parameters of $q,$ and deduce the new site parameters for $f_{i}$
as follows: 
\[
\bm{Q}_{i}\leftarrow\bm{\Sigma}_{h}^{-1}-\bm{Q}_{-i},\quad\bm{r}_{i}\leftarrow\bm{\Sigma}_{h}^{-1}\bm{\mu}_{h}-\bm{r}_{-i}.
\]

EP proceeds by looping over sites, updating each one in turn until
convergence is achieved. In well-behaved cases, one observes empirically
that a small number of complete sweeps through the sites is sufficient
to obtain convergence. However, there is currently no general theory
on the convergence of EP. 

Appendix A gives an algorithmic description of EP, in the more general
case where an exponential family is used for the sites.

\subsection{Approximation of the evidence \label{sub:Evidence-calculation}}

EP also provides an approximation of the normalising constant $Z_{\pi}$
of \eqref{eq:EPdecomposition}, using the same ideas of updating site
approximations through moment matching. To that effect, we rewrite
the EP approximation with normalising constants for each site (assuming
again the prior is Gaussian and does not need to be approximated):
\begin{equation}
q(\bt)=p(\bt)\prod_{i=1}^{n}\frac{f_{i}(\bt)}{C_{i}},\quad f_{i}(\bt)=\exp\left(-\frac{1}{2}\bt^{t}\bm{Q}_{i}\bt+\bm{r}_{i}^{t}\bt\right).\label{eq:q-factorised-normalised}
\end{equation}

Then the update of site $i$ proceeds as before, by adjusting $C_{i}$,
$\bm{r}_{i}$ and $\bm{Q}_{i}$ through moment matching. Simple calculations
\citep[see e.g.][]{Seeger:EPExpFam} lead to the following expressions
for the update of $C_{i}$: 
\begin{equation}
\log\left(C_{i}\right)=\log\left(Z_{h}\right)-\Psi(\bm{r},\bm{Q})+\Psi(\bm{r}_{-i},\bm{Q}_{-i})\label{eq:Ci}
\end{equation}
where $Z_{h}$ is the normalising constant of the hybrid, as defined
in \eqref{eq:EP_moments_Gaussiancase}, $\bm{r}$, $\bm{Q}$ (resp.
$\bm{r}_{-i}$, $\bm{Q}_{-i}$) are the natural parameters of the
current Gaussian approximation $q$ (resp. of $q/f_{i}\propto\prod_{j\neq i}f_{j}$)
and $\Psi(\bm{r},\bm{Q})$ is the log-normalising constant of an unnormalised
Gaussian density:
\[
\Psi(\bm{r},\bm{Q})=\log\left\{ \int\exp\left(-\frac{1}{2}\bm{\bt^{t}Q}\bt+\bm{r}^{\bm{t}}\bt\right)\, d\bt\right\} =-\frac{1}{2}\log\left|\bm{Q}/2\pi\right|+\frac{1}{2}\bm{r}^{t}\bm{Q}\bm{r}.
\]
For each site update, one calculates $C_{i}$ as defined in \eqref{eq:Ci}.
Then, at the end of the algorithm, one may return the following quantity
\[
\sum_{i=1}^{n}\log\left(C_{i}\right)+\Psi(\bm{r},\bm{Q})-\Psi(\bm{r}_{0},\bm{Q}_{0})
\]
as an approximation to the logarithm of the evidence.

\section{EP-ABC: Adapting EP to likelihood-free settings\label{sec:ABC-EP}}

\subsection{Basic principle}

As explained in the introduction, our objective is to approximate
the following ABC posterior 

\begin{equation}
p_{\epsilon}(\bt|\by^{\star})\propto p(\bt)\prod_{i=1}^{n}\left\{ \int p(y_{i}|y_{1:i-1}^{\star},\bt)\I_{\left\{ \Vert s_{i}(y_{i})-s_{i}(y_{i}^{\star})\Vert\leq\varepsilon\right\} }\, dy_{i}\right\} \label{eq:ABCpost2}
\end{equation}
which corresponds to a particular factorisation of the likelihood,
\begin{equation}
p(\by^{\star}|\bt)=\prod_{i=1}^{n}p(y_{i}^{\star}|y_{1:i-1}^{\star},\bt).\label{eq:factor_lik}
\end{equation}
Note that, in full generality, the $y_{i}^{\star}$ may be any type
of ``chunk'' of the observation vector $\by^{\star}$, i.e. the
random variables $y_{i}^{\star}$ may have a different dimension,
or more generally different supports. For simplicity, we assume that
the prior $p(\bt)$ is Gaussian, with natural parameters $\bm{Q}_{0}$
and $\bm{r}_{0}$.

One may interpret \eqref{eq:ABCpost2} as an artificial posterior
distribution, which decomposes into a prior times $n$ likelihood
contributions $l_{i}$, as in \eqref{eq:EPdecomposition}, with 
\[
l_{i}(\bt)=\left\{ \int p(y_{i}|y_{1:i-1}^{\star},\bt)\I_{\left\{ \Vert s_{i}(y_{i})-s_{i}(y_{i}^{\star})\Vert\leq\varepsilon\right\} }\, dy_{i}\right\} .
\]
We have seen that the feasibility of the EP algorithm is determined
by the tractability of the following operation: to compute the two
first moments of a pseudo-posterior, made of a Gaussian prior $N_{d}(\bm{\mu}_{-i},\bm{\Sigma}_{-i})$,
times a single likelihood contribution $l_{i}(\bt)$. This immediately
suggests the following EP-ABC algorithm. We use the EP algorithm,
as described in Algorithm \ref{alg:Generic-EP}, and where the moments
of such a pseudo-posterior are computed using as described in Algorithm
\ref{alg:Computing-the-moments-basic}, that is, as Monte Carlo estimates,
based on simulated pairs $(\bt^{[m]},y_{i}^{[m]})$, where $\bt^{[m]}\sim N_{d}(\bm{\mu}_{-i},\mbox{\ensuremath{\bm{\Sigma}}}_{-i})$,
and $y_{i}^{[m]}|\bt^{[m]}\sim p(y_{i}|y_{1:i-1}^{\star},\bt^{[m]})$. 

\begin{algorithm}[h]
Inputs: $\epsilon$, $\by^{\star}$, $i$, and the moment parameters
$\bm{\mu}_{-i}$, $\bm{\Sigma}_{-i}$ of the Gaussian pseudo-prior
$q_{-i}$. 
\begin{enumerate}
\item Draw $M$ variates $\bt^{[m]}$ from a $N(\bm{\mu}_{-i},\bm{\Sigma}_{-i})$
distribution. 
\item For each $\bt^{[m]}$, draw $y_{i}^{[m]}\sim p(y_{i}|y_{1:i-1}^{\star},\bt^{[m]})$.
\item Compute the empirical moments 
\begin{equation}
M_{acc}=\sum_{m=1}^{M}\I_{\left\{ \Vert s_{i}(y_{i}^{[m]})-s_{i}(y_{i}^{\star})\Vert\leq\varepsilon\right\} },\quad\widehat{\bm{\mu}}_{h}=\frac{\sum_{m=1}^{M}\bt^{[m]}\I_{\left\{ \Vert s_{i}(y_{i}^{[m]})-s_{i}(y_{i}^{\star})\Vert\leq\varepsilon\right\} }}{M_{acc}}\label{eq:Macc}
\end{equation}
\begin{equation}
\widehat{\bm{\Sigma}}_{h}=\frac{\sum_{m=1}^{M}\bt^{[m]}\left\{ \bt^{[m]}\right\} ^{t}\I_{\left\{ \Vert s_{i}(y_{i}^{[m]})-s_{i}(y_{i}^{\star})\Vert\leq\varepsilon\right\} }}{M_{acc}}-\widehat{\bm{\mu}}_{h}\widehat{\bm{\mu}}_{h}^{t}.\label{eq:Sigacc}
\end{equation}

\end{enumerate}
Return $\widehat{Z}_{h}=M_{acc}/M$, $\widehat{\bm{\mu}}_{h}$ and
$\widehat{\bm{\Sigma}}_{h}$. 

\caption{Computing the moments of the hybrid distribution in the likelihood-free
setting, basic algorithm.\foreignlanguage{english}{\label{alg:Computing-the-moments-basic} }}
\end{algorithm}

Since EP-ABC integrates one data-point at a time, it does not suffer
from a curse of dimensionality with respect to $n$: the rejection
rate of Algorithm \ref{alg:Computing-the-moments-basic} corresponds
to a single constraint $\Vert s_{i}(y_{i})-s_{i}(y_{i}^{\star})\Vert\leq\epsilon$,
not $n$ of them, and is therefore likely to be tolerably small even
for small windows $\epsilon$. (Otherwise, in very challenging situations,
one has the liberty to replace Algorithm \ref{alg:Computing-the-moments-basic}
by a more elaborate ABC algorithm.)

The only requirement of EP-ABC is that the factorisation of the likelihood,
\eqref{eq:factor_lik}, is chosen in such a way that simulating from
the model, i.e. $\by\sim p(\by|\bt)$ can be decomposed into a sequence
of steps, where one samples from $p(y_{i}|y_{1:i-1},\bt)$, for $i=1,\ldots,n$.
We shall see in our examples section, see Section \ref{sec:Numerical_examples},
that several important applications of likelihood-free inference fulfil
this requirement. We shall also discuss in Section \ref{sec:Extensions}
how other likelihood-free situations may be accommodated by the EP-ABC
approach.

\subsection{Numerical stability \label{sub:Numerical-stability}}

EP-ABC is a stochastic version of EP, a deterministic algorithm, hence
some care must be taken to ensure numerical stability. We describe
here three strategies towards this aim.

First, to ensure that the stochastic error introduced by each site
update does not vary too much in the course of the algorithm, we adapt
dynamically $M$, the number of simulated points, as follows. For
a given site update, we sample repetitively $M_{0}$ pairs $(\bt^{[m]},y_{i}^{[m]})$,
as described in Algorithm \ref{alg:Computing-the-moments-basic},
until the total number of accepted points exceeds some threshold $M_{\min}$.
Then we compute the moments \eqref{eq:Macc} and \eqref{eq:Sigacc}
based on all the accepted pairs. 

Second, EP-ABC computes a great deal of Monte Carlo estimates, based
on IID (independent and identically distributed) samples, part of
which are Gaussian. Thus, it seems worthwhile to implement variance
reduction techniques that are specific to the Gaussian distribution.
After some investigation, we recommend the following quasi-Monte Carlo
approach. We generate a Halton sequence $\bm{\xi}^{[m]}$ of dimension
$d$, which is a low discrepancy sequence in $[0,1]^{d}$, and take 

\[
\bt^{[m]}=\bm{\mu}_{-i}+\bm{L}_{-i}\Phi^{-1}\left(\bm{\xi}^{[m]}\right),\quad\bm{L}_{-i}\bm{L}_{-i}^{t}=\bm{\Sigma}_{-i}
\]
where $\bm{\mu}_{-i}$, $\bm{\Sigma}_{-i}$ are the moment parameters
corresponding to the natural parameters $\bm{r}_{-i}$, $\bm{Q}_{-i}$,
$\bm{L}_{-i}$ is the Cholesky lower triangle of $\bm{\Sigma}_{-i}$,
and $\Phi^{-1}$ returns a vector that contains the $N(0,1)$ inverse
distribution function of each component of the input vector. We recall
briefly that a low discrepancy sequence in $[0,1]^{d}$ is a deterministic
sequence that spreads more evenly over the hyper-cube $[0,1]^{d}$
than a sample from the uniform distribution would; we refer the readers
to e.g. Chap. 3 of \citet{book:gentle} for a definition of Halton
and other low discrepancy sequences, and the theory of quasi-Monte
Carlo. Rigorously speaking, this quasi-Monte Carlo version of EP-ABC
is a hybrid between Monte Carlo and quasi-Monte Carlo, because the
$y_{i}^{[m]}$ are still generated using standard Monte Carlo. However,
we do observe a dramatic improvement when using this quasi-Monte Carlo
approach. An additional advantage is that one may save some computational
time by generating once and for all a very large sequence of $\Phi^{-1}\left(\bm{\xi}^{[m]}\right)$
vectors, and store it in memory for all subsequent runs of EP-ABC.

The third measure we may take is to slow down the progression of the
algorithm such as to increase stability, by conservatively updating
the parameters of the approximation in Step 3 of Algorithm \ref{alg:Generic-EP},
that is, $\bl_{i}\leftarrow\alpha(\bl_{h}-\bl_{-i})+(1-\alpha)\bl_{i}$.
Standard EP is the special case with $\alpha=1$. Updates of this
type are suggested in \citet{minka2004power}.

In our experiments, we found that the two first strategies improve
performance very significantly (in the sense of reducing Monte Carlo
variability over repeated runs), and that the third strategy is sometimes
useful, for example in our reaction time example, see Section \ref{sub:Reaction-times}.

\subsection{Evidence approximation}

In this section, we consider the special case $s_{i}(y_{i})=y_{i}$,
and we normalise the ABC posterior \eqref{eq:ABCpost2} as follows:
\begin{equation}
p_{\epsilon}(\bt|\by^{\star})=\frac{1}{p_{\epsilon}(\by^{\star})}p(\bt)\prod_{i=1}^{n}\left\{ \int p(y_{i}|y_{1:i-1}^{\star},\bt)\frac{\I_{\left\{ \Vert y_{i}-y_{i}^{\star}\Vert\leq\varepsilon\right\} }}{v_{i}(\epsilon)}\, dy_{i}\right\} ,\label{eq:ABCpost_normalise}
\end{equation}
where $v_{i}(\epsilon)$ is the normalising constant of the uniform
distribution with respect to the ball of centre $y_{i}^{\star}$,
radius $\epsilon$, and norm $\Vert\cdot\Vert$. For the Euclidean
norm, and assuming that the $y_{i}$'s have the same dimension $d_{y}$,
one has: $v_{i}(\epsilon)=v_{i}(1)\epsilon^{d_{y}}$, with $v_{i}(1)=\pi^{d_{y}/2}/\Gamma(d_{y}/2+1)$;
e.g. $v_{i}(1)=2$ if $d_{y}=1$, $v_{i}(1)=\pi$ if $d_{y}=2$. 

\citet{wilkinson2008approximate} shows that a standard ABC posterior
such as \eqref{eq:ABCpost} can be interpreted as the posterior distribution
of a new model, where the summary statistics are corrupted with a
uniformly-distributed noise (assuming these summary statistics are
sufficient). The expression above indicates that this interpretation
also holds for this type of summary-less ABC posterior, except that
the artificial model is now such all the random variables $y_{i}$
are corrupted with noise (conditional on $y_{1:i-1}^{\star}$). 

The expression above also raises an important point regarding the
approximation of the evidence. In \eqref{eq:ABCpost_normalise}, the
normalising constant $p_{\epsilon}(\by^{\star})$ is the evidence
of the corrupted model, which converges to the evidence $p(\by^{\star})$
of the actual model as $\epsilon\rightarrow0$. On the other hand,
EP-ABC targets \eqref{eq:ABCpost2}, and, in particular, see Section
\ref{sub:Evidence-calculation}, produces an EP approximation of its
normalising constant, which is $p_{\epsilon}(\by^{\star})\prod_{i=1}^{n}v_{i}(\epsilon)$.
Thus, one needs to divide this EP approximation by $\prod_{i=1}^{n}v_{i}(\epsilon)$
in order recover an approximation of $p_{\epsilon}(\by^{\star})$.
We found in our simulations that, when properly normalised as we have
just described, the approximation of the evidence provided by EP-ABC
is particularly accurate, see Section \ref{sec:Numerical_examples}.
In contrast, standard ABC based on summary statistics cannot provide
an approximation of the evidence, as explained in the Introduction.

\section{Speeding up EP-ABC in the IID case\label{sec:IID_case}}

Typically, the main computational bottleneck of EP-ABC, or other types
of ABC algorithms, is simulating pseudo data-points from the model.
In this section, we explain how these simulations may be recycled
throughout the iterations in the IID (independent and identically
distributed) case, so as to significantly reduce the overall computational
cost of EP-ABC. 

Our recycling scheme is based on a straightforward importance sampling
strategy. Consider an IID model, with likelihood $p(\by^{\star}|\bt)=\prod_{i=1}^{n}p(y_{i}^{\star}|\bt)$.
Also, for the sake of simplicity, take $s_{i}(y_{i})=y_{i}$. Assume
that, for a certain site $i$, pairs $(\bt^{[m]},y^{[m]})$ are generated
from $q_{-i}(\bt)p(y|\bt)$, as described in Algorithm \ref{alg:Computing-the-moments-basic}.
We have removed the subscript $i$ in both $y^{[m]}$ and $p(y|\bt)$,
to highlight the fact that the generative process of the data-points
is the same for all the sites. The next update, for site $i+1$, requires
computing moments with respect to $q_{-(i+1)}(\bt)p(y|\bt)\I\left\{ \Vert y-y_{i+1}^{\star}\Vert\leq\epsilon\right\} $.
Thus, we may recycle the simulations of the previous site by assigning
to each pair $(\bt^{[m]},y^{[m]})$ the importance sampling weight:
\[
w_{i+1}^{[m]}=\frac{q_{-(i+1)}\left(\bt^{[m]}\right)}{q_{-i}\left(\bt^{[m]}\right)}\times\I\left\{ \Vert y^{[m]}-y_{i+1}^{\star}\Vert\leq\epsilon\right\} 
\]
and compute the corresponding weighted averages. 

Obviously, this step may also be applied to the subsequent sites,
$i+2,i+3,\ldots$, until one reaches a stage when the weighted sample
is too degenerated. When this happens, ``fresh'' simulations may
be generated from the current site. Algorithm \ref{alg:Computing-the-moments-recycle-IID}
describes more precisely this recycling strategy. To detect weight
degeneracy, we use the standard ESS (Effective Sample Size) criterion
of \citet{KongLiuWong}: we regenerate when the ESS is smaller that
some threshold $\mathrm{ESS}_{\min}$. 

\begin{algorithm}[h]
Inputs: $i,$ $\epsilon$, current weighted sample $(\bt^{[m]},y^{[m]})_{m=1}^{M}$,
moment parameters $\tilde{\bm{\mu}}$ and $\tilde{\bm{\Sigma}}$ (resp.
$\bm{\mu}_{-i}$ and $\bm{\Sigma}_{-i}$) that correspond to the site
where data were re-generated for the last time (resp. that correspond
to the Gaussian approximation $\prod_{j\neq i}q_{j}(\bt)$). 
\begin{enumerate}
\item Compute the importance sampling weights 
\[
w_{i}^{[m]}=\frac{N(\bt^{[m]};\bm{\mu}_{-i},\bm{\Sigma}_{-i})}{N(\bt^{[m]};\tilde{\bm{\mu}},\tilde{\bm{\Sigma}})}\times\I_{\left\{ \Vert y^{[m]}-y_{i}^{\star}\Vert\leq\epsilon\right\} }
\]
where $N(\bt;\bm{\mu},\bm{\text{\ensuremath{\Sigma}}})$ stands for
the Gaussian $N(\bm{\mu},\bm{\Sigma})$ probability density evaluated
at $\bt$, $\bm{}$and the effective sample size: 
\[
\mathrm{ESS}=\frac{\left(\sum_{m=1}^{M}w_{i}^{[m]}\right)^{2}}{\sum_{m=1}^{M}\left(w_{i}^{[m]}\right)^{2}}.
\]

\item If $\mathrm{ESS}<\mathrm{ESS}_{\min}$, replace $(\bt^{[m]},y^{[m]})_{m=1}^{M}$
by $M$ IID draws from $q_{-i}(\bt)p(y|\bt)$, set $w_{i}^{[m]}=\I_{\left\{ \Vert y^{[m]}-y_{i}^{\star}\Vert\leq\epsilon\right\} }$,
and $\tilde{\bm{\mu}}=\bm{\mu}_{-i}$, $\tilde{\bm{\Sigma}}=\bm{\Sigma}_{-i}$. 
\item Compute the following importance sampling estimates: 
\[
\widehat{Z}_{h}=\frac{1}{M}\sum_{m=1}^{M}w_{i}^{[m]},\quad\widehat{\bm{\mu}}_{h}=\frac{\sum_{m=1}^{M}w_{i}^{[m]}\bm{\theta}^{[m]}}{\widehat{Z}_{h}}
\]
and 
\[
\widehat{\bm{\Sigma}}_{h}=\frac{\sum_{m=1}^{M}w_{i}^{[m]}\bm{\theta}^{[m]}\left\{ \bm{\theta}^{[m]}\right\} ^{t}}{\widehat{Z}_{h}}-\widehat{\bm{\mu}}_{h}\widehat{\bm{\mu}}_{h}^{t}.
\]

\end{enumerate}
Return $(\bt^{[m]},y^{[m]})_{m=1}^{M}$, $\tilde{\bm{\mu}}$, $\tilde{\bm{\Sigma}}$,
$\widehat{Z}_{h}$, $\widehat{\bm{\mu}}_{h}$ and $\widehat{\bm{\Sigma}}_{h}$. 

\caption{\selectlanguage{english}%
Computing the moments of the hybrid distribution in the likelihood-free
setting, recycling scheme for IID models.\label{alg:Computing-the-moments-recycle-IID} \selectlanguage{british}%
}
\end{algorithm}

The slower the EP approximation evolves, the less often regenerating
the pseudo data-points is necessary, so that as the approximation
gradually stabilises, we do not need to draw any new samples any more.
Since EP slows down rapidly during the first two passes, most of the
computational effort will be devoted to the early phase, and additional
passes through the data will come essentially for free. 

In non-IID cases several options are still available. For some models
the data may come in blocks, each block made of IID data-points (think
for example of a linear model with discrete predictors). We can apply
the strategy outlined above in a block-wise manner (see the reaction
times example, section \ref{sub:Reaction-times}). In other models
there may be an easy transformation of the samples for data-point
$i$ such that they become samples for data-point $j\neq i$, or one
may be able to reuse part of the simulation.

\section{Case studies\label{sec:Numerical_examples}}

\subsection{General methodology}

In each scenario, we apply the following approach. In a first step,
we run the EP-ABC algorithm. We may run the algorithm several times,
to evaluate the Monte Carlo variability of the output, and we may
also run it for different values of $\varepsilon$, in order to assess
the sensitivity to this approximation parameter. We use the first
run to determine how many passes (complete cycles through the $n$
sites) are necessary to reach convergence. A simple way to monitor
convergence is to plot the evolution of the expectation (or the smallest
eigenvalue of the covariance matrix) of the current Gaussian approximation
$q$ along the site updates. Note however that, in our experience,
it is quite safe to simply fix the number of complete passes to a
small number like $4$. Finally, note that in each example, we could
take $s_{i}(y_{i})=y_{i}$, so the point of determining appropriate
local summary statistics is not discussed. 

In a second step, we run alternative algorithms, that is, either an
exact (but typically expensive) MCMC algorithm, or an ABC algorithm,
based on some set of summary statistics. The ABC algorithm we implement
is a Gaussian random walk version of the MCMC-ABC algorithm of \citet{Marjoram:MCMCWithoutLikelihood}.
This algorithm targets a standard ABC approximation, i.e. \eqref{eq:ABCpost},
that corresponds to a single constraint $\left\{ \left\Vert \bm{s}(\by)-\bm{s}(\by^{\star})\right\Vert \leq\epsilon\right\} $,
for some vector of summary statistics $\bm{s}$, and some $\epsilon$;
the specific choices of $\bm{s}$ and $\epsilon$ are discussed for
each application. We calibrate the tuning parameters of these MCMC
algorithms using the information provided by the first step: we use
as a starting point for the MCMC chain the expectation of the approximated
posterior distribution provided by the EP-ABC algorithm, random walk
scales are taken to be some fraction of the square root of the approximated
posterior variances, and so on. This makes our comparisons particularly
unfavourable to EP-ABC. Despite this, we find consistently that the
EP-ABC algorithm is faster by several orders of magnitude, and leads
to smaller overall approximation errors. We report computational loads
both in terms of CPU time (e.g. 30 seconds) and in terms of the number
of simulations of replicate data-points $y_{i}$. The latter should
be typically the bottleneck of the computation. 

All the computations were performed on a standard desktop PC in Matlab;
programs are available from the first author's web page.

\subsection{First example: Alpha-stable Models\label{sub:First-example:-Alpha-stable}}

Alpha-stable distributions are useful in areas (e.g. Finance) concerned
with noise terms that may be skewed, may have heavy tails and an infinite
variance. A univariate alpha-stable distribution does not admit a
close-form expression for its density, buy may be specified through
its characteristic function

\[
\Phi_{X}(t)=\begin{cases}
\exp\left[i\delta t-\gamma^{\alpha}\left|t\right|^{\alpha}\left\{ 1+i\beta\left(\tan\frac{\pi\alpha}{2}\right)\mathrm{sgn}(t)\left(\left|\gamma t\right|^{1-\alpha}-1\right)\right\} \right] & \alpha\neq1\\
\exp\left[i\delta t-\gamma\left|t\right|\left\{ 1+i\beta\frac{2}{\pi}\mathrm{sgn}(t)\log\left|\gamma t\right|\right\} \right] & \alpha=1
\end{cases}
\]
where $\alpha$ determines the tails, $0<\alpha\leq2$, $\beta$ determines
skewness, $-1<\beta<1$, and $\gamma>0$ and $\delta$ are respectively
scale and location parameters; see \citet[Chap. 1]{nolan:2012} for
a general introduction to stable distributions. 

\citet{Peters-alphastable} consider a model of $n$ i.i.d. observations
$y_{i}$, $i=1,\ldots,n$ from a univariate alpha-stable distribution,
and propose to use the ABC approach to infer the parameters. Likelihood-free
inference is appealing in this context, because sampling from an alpha-stable
distribution is fast \citep[using e.g. the algorithm of ][]{chambers1976method},
while computing its density is cumbersome.

Trying EP-ABC on this example is particularly interesting for the
following reasons: (a) \citet{Peters-alphastable} show that choosing
a reasonable set of summary statistics for this problem is difficult,
and that several natural choices lead to strong biases; and (b) since
alpha-stable distributions are very heavy-tailed, the posterior distribution
may be heavy-tailed as well, which seems a challenging problem for
a method based on a Gaussian approximation such as EP-ABC. 

Our data consist of $n=1264$ rescaled log-returns, $y_{t}=100*\log(z_{t}/z_{t-1})$,
computed from daily exchange rates $z_{t}$ of AUD (Australian Dollar)
recorded in GBP (British Pound) between 1 January 2005 and 1 December
2010. (These data are publicly available on the Bank of England web-site.)
We take $\bt=\left(\Phi^{-1}(\alpha/2),\Phi^{-1}\left((\beta+1)/2\right),\log\gamma,\delta\right)$
where $\Phi$ is the $N(0,1)$ cumulative distribution function, and
we set the prior to $N\left(0_{4},\mathrm{diag}(1,1,10,10)\right)$.
Note however that our results are expressed in terms of the initial
parametrisation $\alpha$, $\beta$, $\gamma$ and $\delta$; i.e.
for each parameter we report the approximate marginal posterior distribution
obtained through the appropriate variable transform of the Gaussian
approximation produced by EP-ABC. We run the EP-ABC algorithm (recycling
version, as model is IID, see Section \ref{sec:IID_case}), with $\epsilon=0.1$,
$M=8\times10^{6}$, $\mathrm{ESS}_{\min}=2\times10^{4}$, and $\Vert\cdot\Vert$
set to the Euclidean norm in $\mathbb{R}$ (i.e. the $n$ constraints
in \eqref{eq:ABCpost2} simplify to $|y_{i}-y_{i}^{\star}|\leq\varepsilon$).
Variations over ten runs are negligible. Average CPU time for one
run is 39 minutes, and average number of simulated data-points over
the course of the algorithm, is $4\times10^{8}$. 

\begin{figure}
\centering{}\includegraphics[scale=0.7]{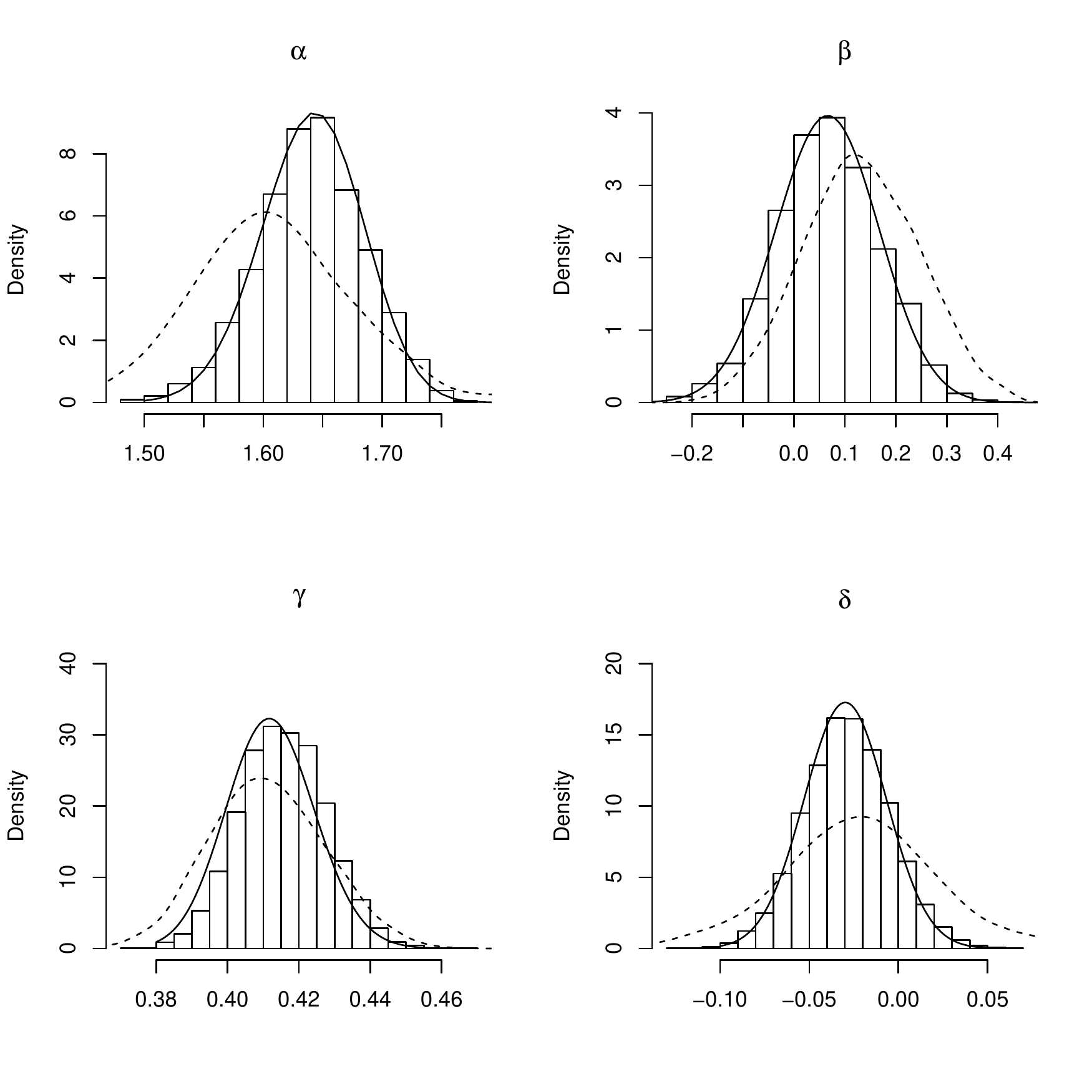}\caption{\selectlanguage{english}%
\label{fig:Results-for-alpha-stable}Marginal posterior distributions
of $\alpha$, $\beta,$ $\gamma$ and $\delta$ for alpha-stable model:
MCMC output from the exact algorithm (histograms), approximate posteriors
provided by first run of EP-ABC (solid line), kernel density estimates
computed from MCMC-ABC sample based on summary statistic proposed
by \citet{Peters-alphastable} (dashed line). \selectlanguage{british}%
}
\end{figure}

We first compare these results with the output of an exact random-walk
Hastings-Metropolis algorithm, which relies on the evaluation of an
alpha-stable probability density function for each data-point (using
numerical integration). Because of this, this algorithm is very expensive.
We ran the exact algorithm for about 60 hours ($2\times10^{5}$ iterations).
One sees in Figure \ref{fig:Results-for-alpha-stable} that the difference
between EP-ABC and the exact algorithm is negligible.

We then compare these results with those obtained by MCMC-ABC, for
the set of summary statistics which performs best among those discussed
by \citet[see $S_1$ in Section 3.1]{Peters-alphastable}. We run $2\times10^{7}$
iterations of this sampler, which leads to about $50$ times more
simulations from an univariate alpha-stable distribution than in the
EP-ABC runs above. Through pilot runs, we decided to set $\epsilon=0.03$,
which seems to be as small as possible, subject to having a reasonable
acceptance rate ($2\times10^{-3}$) for this computational budget.
In Figure \ref{fig:Results-for-alpha-stable}, we see that the posterior
output from this MCMC-ABC exercise is not as good an approximation
as the output of EP-ABC. As explained in the previous section, we
have set the starting point of the MCMC-ABC chain to the posterior
mode. If initialised from some other point, the sampler typically
takes a much longer time to reach convergence, because the acceptance
rate is significantly lower in regions far from the posterior mode. 

Finally, we also use EP-ABC, with the same settings as above, e.g.
$\epsilon=0.1$, in order to approximate the evidence of the model
above ($-1385.8$) and two alternative models, namely a symmetric
alpha-stable model, where $\beta$ is set to $0$ ($-1383.8)$, and
a Student model ($-1383.6$), with 3 parameters (scale $\gamma$,
position $\delta$, degrees of freedom $\nu$, and a Gaussian prior
$N(0_{3},\mathrm{diag}(10,10,10))$ for $\bt=(\log\nu,\gamma,\delta)$).
(Standard deviation over repeated runs is below $0.1$.) One sees
that there no strong evidence of skewness in the data, and that the
Student distribution and a symmetric alpha-stable distribution seem
to fit equally well the data. We obtained the same value ($-1383.6$)
for the evidence of the Student model when using the generalised harmonic
mean estimator \citep{gelfand1994bayesian} based on a very long chain
of an exact MCMC algorithm. For both alpha-stable models, this approach
proved to be too expensive to allow for a reliable comparison.

\subsection{Second example: Lotka-Volterra models}

The stochastic Lotka-Volterra process describes the evolution of two
species $Y_{1}$ (prey) and $Y_{2}$ (predator) through the reaction
equations: 
\begin{eqnarray*}
Y_{1} & \stackrel{r_{1}}{\rightarrow} & 2Y_{1}\\
Y_{1}+Y_{2} & \stackrel{r_{2}}{\rightarrow} & 2Y_{2}\\
Y_{2} & \stackrel{r_{3}}{\rightarrow} & \emptyset.
\end{eqnarray*}
This chemical notation means that, in an interval $[t,t+dt]$, the
probability that one prey is replaced by two preys is $r_{1}d_{t}$,
and so on. Typically, the observed data $\by^{\star}=(y_{1},\ldots,y_{n})$
are made of $n$ vectors $y_{i}^{\star}=(y_{i,1}^{\star},y_{i,2}^{\star})$
in $\mathbb{N}^{2}$, which correspond to the population levels at
integer times. We take $\bt=\left(\log r_{1},\log r_{2},\log r_{3}\right)$.
This model is Markov, $p(y_{i}^{\star}|y_{1:i-1}^{\star},\bt)=p(y_{i}^{\star}|y_{i-1}^{\star},\bt)$
for $i>1$, and one can efficiently simulate from $p(y_{i}^{\star}|y_{i-1}^{\star},\bt)$
using \citet{Gillespie:ExactStochasticSimulation}'s algorithm. On
the other hand, the density $p(y_{i}^{\star}|y_{i-1}^{\star},\bt)$
is intractable. This makes this model a clear candidate both for ABC,
as noted by \citet{Toni:ABCDynamicalSystems}, and for EP-ABC. \citet{boys2008bayesian}
show that MCMC remains feasible for this model, but in certain scenarios
the proposed schemes are particularly inefficient, as noted also by
\citet[Chap. 4]{Holenstein-phd}.

\begin{figure}
\centering{}\includegraphics[scale=0.6]{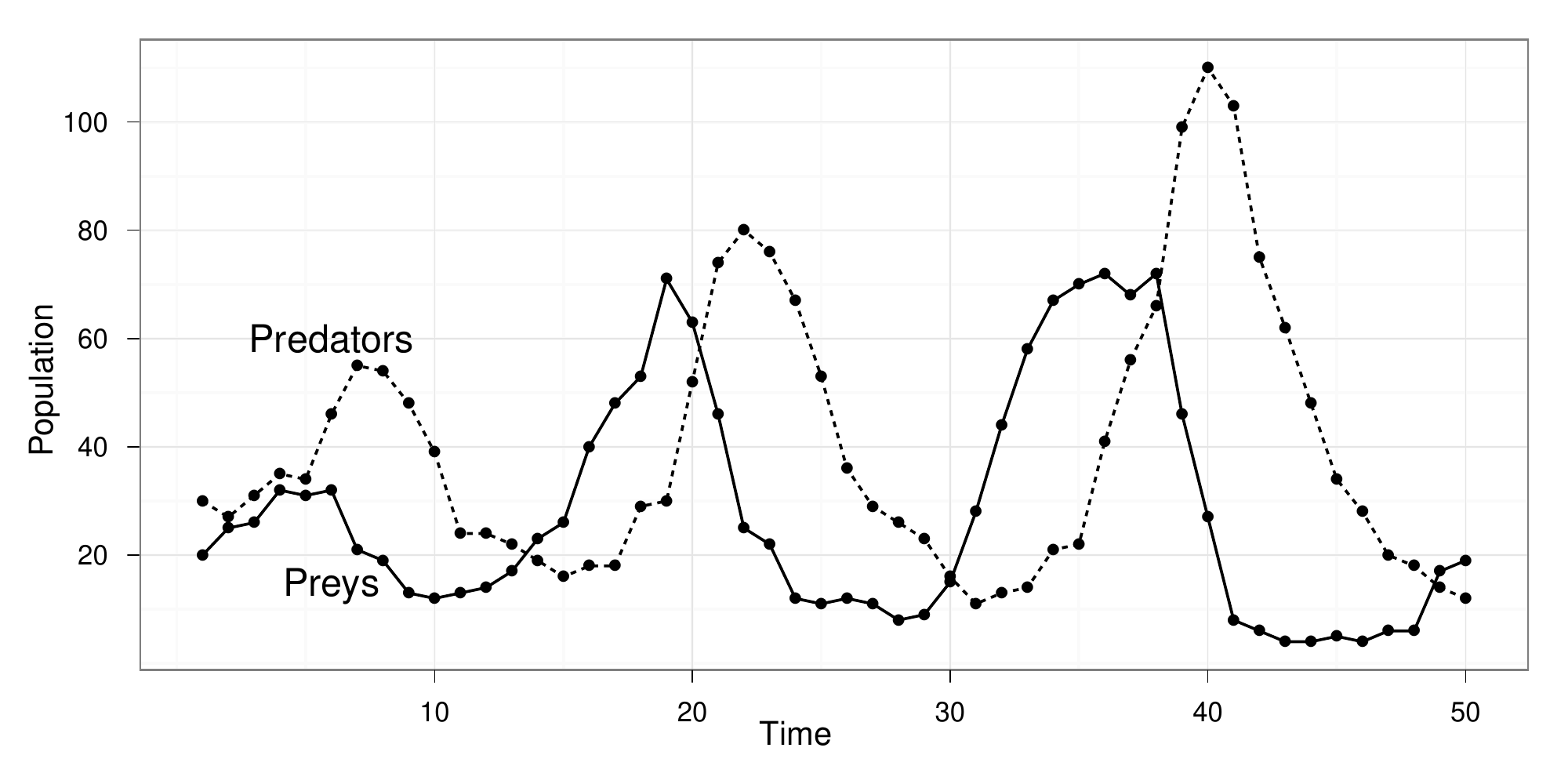}\caption{\selectlanguage{english}%
Lotka-Volterra example: simulated dataset\label{fig:Lokta-Volterra-example:-simulated}\selectlanguage{british}%
}
\end{figure}

Following the aforementioned papers, we consider a simulated dataset,
corresponding to rates $r_{1}=0.4$, $r_{2}=0.01$, $r_{3}=0.3$,
initial population values $y_{0,1}^{\star}=20$, $y_{0,2}^{\star}=30$$ $
and $n=50$; see Figure \ref{fig:Lokta-Volterra-example:-simulated}.
Since the observed data are integer-valued, we use the supremum norm
in \eqref{eq:ABCpost2}, and an integer-valued $\epsilon$; this is
equivalent to imposing simultaneously the $2n$ constraints $\left|y_{i,1}-y_{i,1}^{\star}\right|\leq\epsilon$
and $\left|y_{i,2}-y_{i,2}^{\star}\right|\leq\epsilon$ in the ABC
posterior. 

\begin{figure}
\centering{}\includegraphics[scale=0.6]{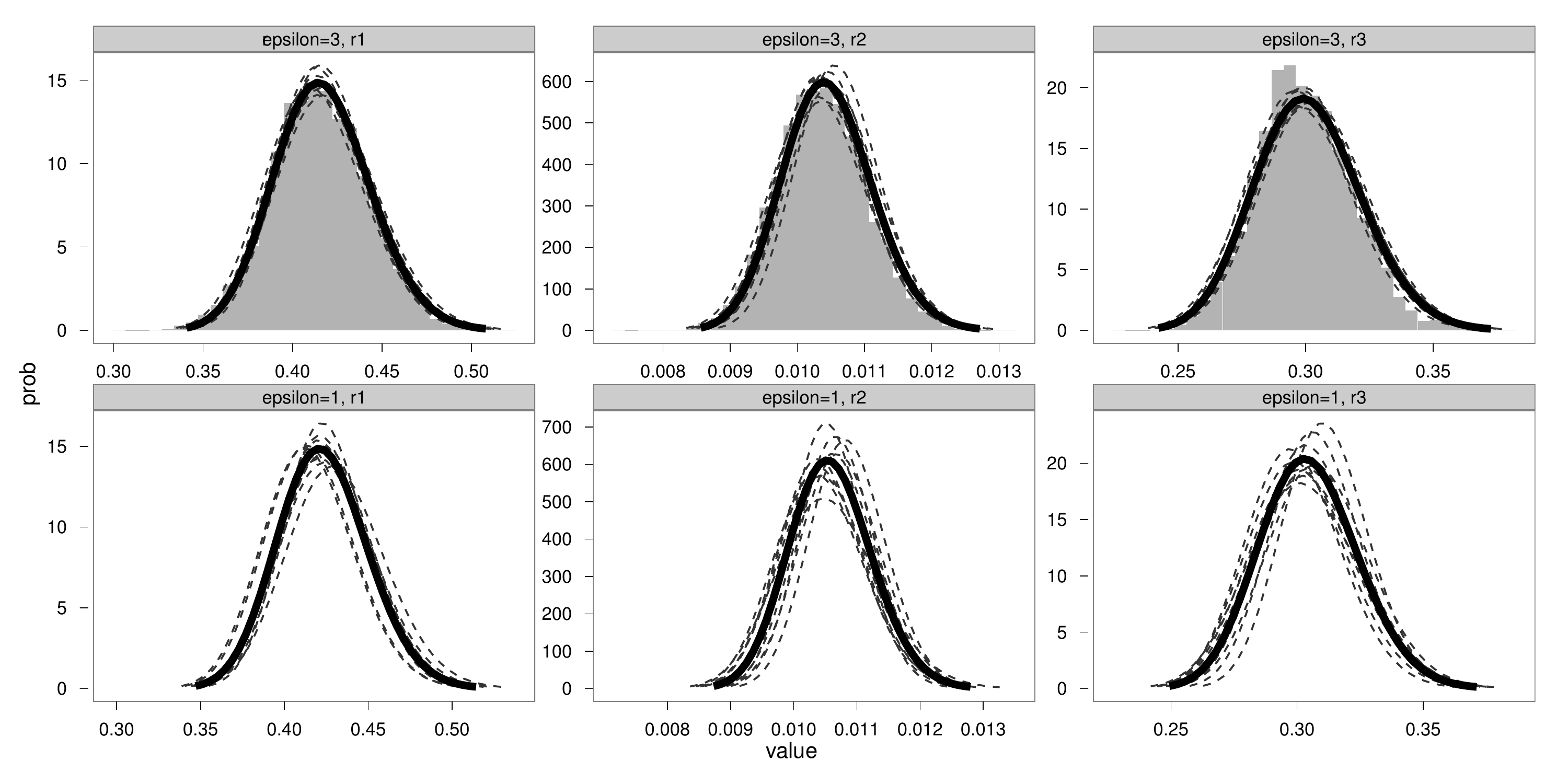}\caption{\selectlanguage{english}%
Lotka-Volterra example: marginal posterior densities of rates $r_{1}$,
$r_{2}$, $r_{3}$, obtained from PMCMC algorithm (histograms), and
from ABC-EP, for $\epsilon=3$ (top) and $\epsilon=1$ (bottom); PMCMC
results for $\epsilon=1$ could not be obtained in a reasonable time.
The solid lines correspond to the average over 10 runs of the moment
parameters of the Gaussian approximation, while the dashed lines correspond
to the 10 different runs. \label{fig:Lokta-Volterra-example:-marginal}\selectlanguage{british}%
}
\end{figure}

First, we run EP-ABC (standard version) with $M_{\min}=4000$, and
for both $\epsilon=3$ and $\epsilon=1$. We find that a single pass
over the data is sufficient to reach convergence. For $\epsilon=3$
(resp. $\epsilon=1$), CPU time for each run is 2.5 minutes (resp.
25 minutes), and number of simulated transitions $p(y_{i}|y_{i-1}^{\star},\bt)$
is about $10^{7}$ (resp. $9\times10^{7}$); marginal posteriors obtained
through EP-ABC are reported in Figure \ref{fig:Lokta-Volterra-example:-marginal}. 

When applying ABC to this model, \citet{Toni:ABCDynamicalSystems}
uses as a pseudo-distance between the actual data $\by^{\star}$ and
the simulated data $\by$ the sum of of squared errors. In \citet{wilkinson2008approximate}'s
perspective discussed in Section \ref{sec:IID_case}, this is equivalent
to considering a state-space model where the latent process is the
Lotka-Volterra process described above, and the observation process
is the same process, but corrupted with Gaussian noise (provided the
indicator function $\I\left\{ \sum(y_{i}-y_{i}^{\star})^{2}\leq\varepsilon\right\} $
in the ABC posterior is replaced by the kernel function $\exp\left\{ -\sum(y_{i}-y_{i}^{\star})^{2}/\varepsilon\right\} $).
Thus, instead of a standard ABC algorithm, it seems more efficient
to resort to a MCMC sampler specifically designed for state-space
models in order to simulate from the ABC approximation of \citet{Toni:ABCDynamicalSystems}.
Following \citet[Chap. 4]{Holenstein-phd}, we consider a Gaussian
random walk version of the marginal PMCMC sampler. This algorithm
is a particular instance of the state of the art PMCMC framework of
\citet{PMCMC}, which is based on the idea of approximating the marginal
likelihood of the data by running a particle filter of size $N$ at
each iteration of the MCMC sampler. The big advantage of PMCMC in
this situation (comparatively to other MCMC approaches for state-space
models), is that it does not requires a tractable probability density
for the Markov transition of the state-space model.

In Figure \ref{fig:Lokta-Volterra-example:-marginal}, we report the
posterior output obtained from this sampler, run for about $2\times10^{5}$
iterations and $N=1000$ particles (2 days in CPU time, $10^{10}$
simulated transitions $p(y_{i}|y_{i-1}^{\star},\bt)$), with random
walk scales set to obtain approximatively a $25\%$ acceptance rate.
These plots correspond to $\epsilon=3$, and a state-space model with
an uniformly distributed observation noise. In Figure \ref{fig:Lokta-Volterra-example:-marginal},
one detects practically no difference between PMCMC and EP-ABC with
$\epsilon=3$ (black lines), although the CPU time of the latter was
about 1500 smaller.

The difference between the two EP-ABC approximations (corresponding
to $\epsilon=1$ and $\epsilon=3$) is a bit more noticeable. Presumably,
the EP-ABC approximation corresponding to $\epsilon=1$ is slightly
closer to the true posterior. We did not manage however to obtain
reliable results from our PMCMC sampler and $\epsilon=1$ in a reasonable
time.

\subsection{Third example: Race models of reaction times\label{sub:Reaction-times}}

Reaction time models seek to describe the decision behaviour of (human
or animal) subjects in a choice task \citep{Luce:ResponseTimes,Meyer:ModernMentalChronometry,Ratcliff:TheoryMemRetrieval}.
In the typical experiment, subjects view a stimulus, and must choose
an appropriate response. For example, the stimulus might be a set
of moving points, and the subject must decide whether the points move
to the left or to the right. 

Assuming that the subject may choose between $k$ alternatives, one
observes independent pairs, $y_{i}=(d_{i},r_{i})$, where $d_{i}\in\{1,\ldots,k\}$
is the chosen alternative, and $r_{i}\geq0$ is the measured reaction
time. For convenience, we drop for now the index $i$ in order to
describe the random distribution of the pair $(d,r)$. 

\begin{figure}
\begin{centering}
\includegraphics[scale=0.5]{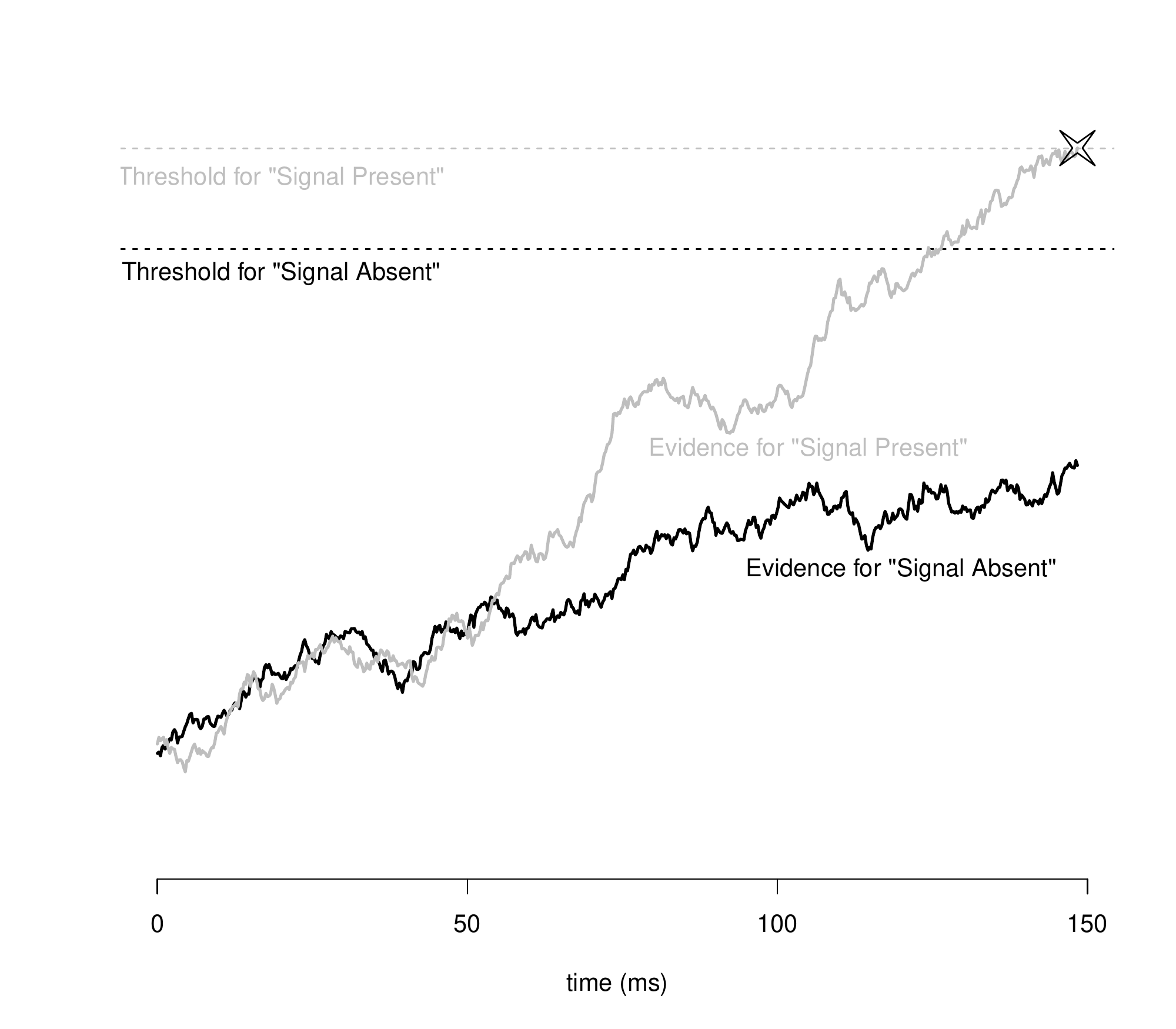}
\par\end{centering}

\caption{A model of reaction time in a choice task. The subject has to choose
between $k$ responses (here, ``Signal present'' and ``Signal absent'')
and information about each accumulates over time in the form of evidence
in favour of one and the other. Because of noise in the neural system
``evidence'' follows a random walk over time. A decision is reached
when the evidence for one option reaches a threshold (dashed lines).
The decision time in this example is denoted by the star: here the
subject decides for 'B' after about 150ms. The fact that the thresholds
are different for ``Signal Present'' and ``Signal Absent'' capture
decisional bias: in general, for the same level of information, the
subject favours option ``Signal Absent''. \label{fig:RTmodel_illustration}}
\end{figure}

Reaction time models assume that the brain processes information progressively,
and that a decision is reached when a sufficient amount of information
has been accumulated. In the model we use here (a variant of typical
models found in e.g. \citealp{RatcliffMcKoon:DiffusionDecisionModelReview,Bogaczetal:BioInspiredModelChoice})
$k$ parallel integrators represent the evidence $e_{1}\left(t\right),\ldots,e_{k}\left(t\right)$
in favour of each of the $k$ alternatives. The model is illustrated
on Figure \ref{fig:RTmodel_illustration}. The first accumulator to
reach its boundary $b_{j}$ wins the race and determines which response
the subject will make. Each accumulator undergoes a Wiener process
with drift:

\[
\tau de_{j}(t)=m_{j}dt+dW_{t}^{j}
\]
where the $m_{j}$'s are the drift parameters, the $W_{t}^{j}$'s
are $k$ independent Wiener processes; and $\tau$ is a fixed time
scale, \textbf{$\tau=5ms$}. The measured reaction time is corrupted
by a uniformly-distributed noise $r_{nd}$, representing the ``non-decisional
time'' \citep{RatcliffMcKoon:DiffusionDecisionModelReview}, i.e.
the time the subject needs to execute the decision (prepare a motor
command, press an answer key, ...). This model is summarised by the
following equations: 
\begin{eqnarray*}
r & = & r_{d}+r_{nd},\quad r_{nd}\sim\mathrm{U}[a,b],\\
r_{d} & = & \min_{j}\inf_{t}\left\{ t:\, e_{j}(t)=b_{j}\right\} ,\\
d & = & \arg\min_{j}\inf_{t}\left\{ t:\, e_{j}(t)=b_{j}\right\} .
\end{eqnarray*}
(We fix $a$ and $b$ to $a=100ms$, $b=200ms$, credible values from
\citet{RatcliffMcKoon:DiffusionDecisionModelReview}) 

The model above captures the essential ideas of reaction time modelling,
but it remains too basic for experimental data. We now describe several
important extensions. First, a better fit is obtained if the boundaries
are allowed to vary randomly from trial to trial (as in \citealp{Ratcliff:TheoryMemRetrieval}):
we assume that $b_{j}=c_{j}+\tau$, where $\tau\sim N(0,e^{s}),$
and $s$ is a parameter to be estimated. Second, a mechanism is needed
to ensure that the reaction times cannot be too large: we assume that
if no boundary has been reached after 1 second, information accumulation
stops and the highest accumulator determines the decision. Finally,
one needs to account for lapses \citep{WichmannHill:PsychometricFunctionI}:
on certain trials, subjects simply fail to pay attention to the stimuli
and respond more or less at random. We account for this phenomenon
by having a lapse probability of 5\%. In case a lapse occurs, $r_{d}$
becomes uniformly distributed between 0 and 800 ms and the response
is chosen between the alternatives with equal probability. Clearly,
although this generalised model remains amenable to simulation, the
corresponding likelihood is intractable. 

\begin{figure}
\subfloat[\selectlanguage{english}%
Probability of answering ``Signal present'' as a function of relative
target contrast in a detection experiment, at three different positions
of the target (data from one subject). Filled dots represent raw data,
the grey curves are the result of fitting a binomial GLM with Cauchit
link function. The light grey band is a 95\% pointwise, asymptotic
confidence band on the fit obtained from the observed Fisher information.
As expected in such experiments, the probability of detecting the
target increases with relative target contrast.\selectlanguage{british}%
]{\begin{centering}
\includegraphics[scale=0.5]{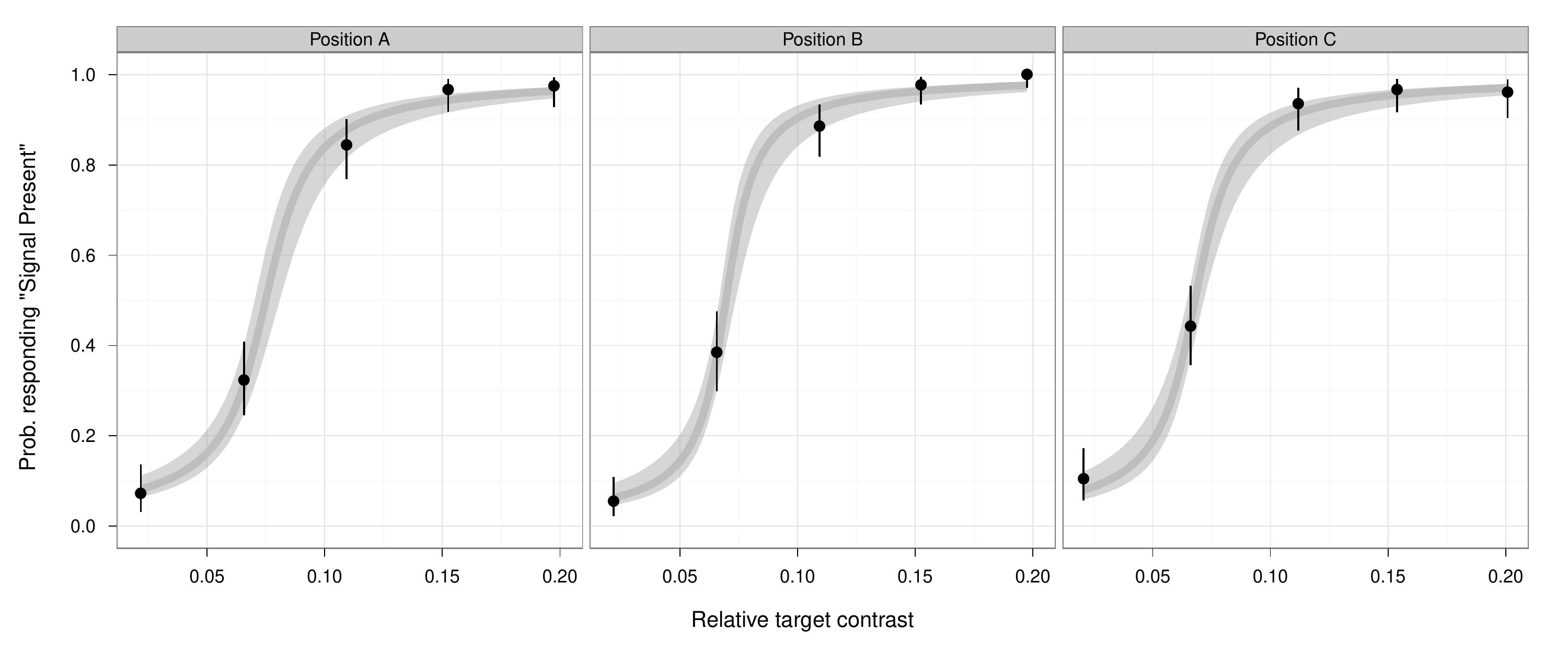}
\par\end{centering}

}

\subfloat[Reaction time distributions conditional on target contrast, target
position, and response. The semi-transparent dots represent reaction
times for individual trials. Horizontal jitter was added to aid visualisation.
The lines represent linear quantile regressions for the 30\%, 50\%
and 70\% quantiles. ]{\begin{centering}
\includegraphics[scale=0.5]{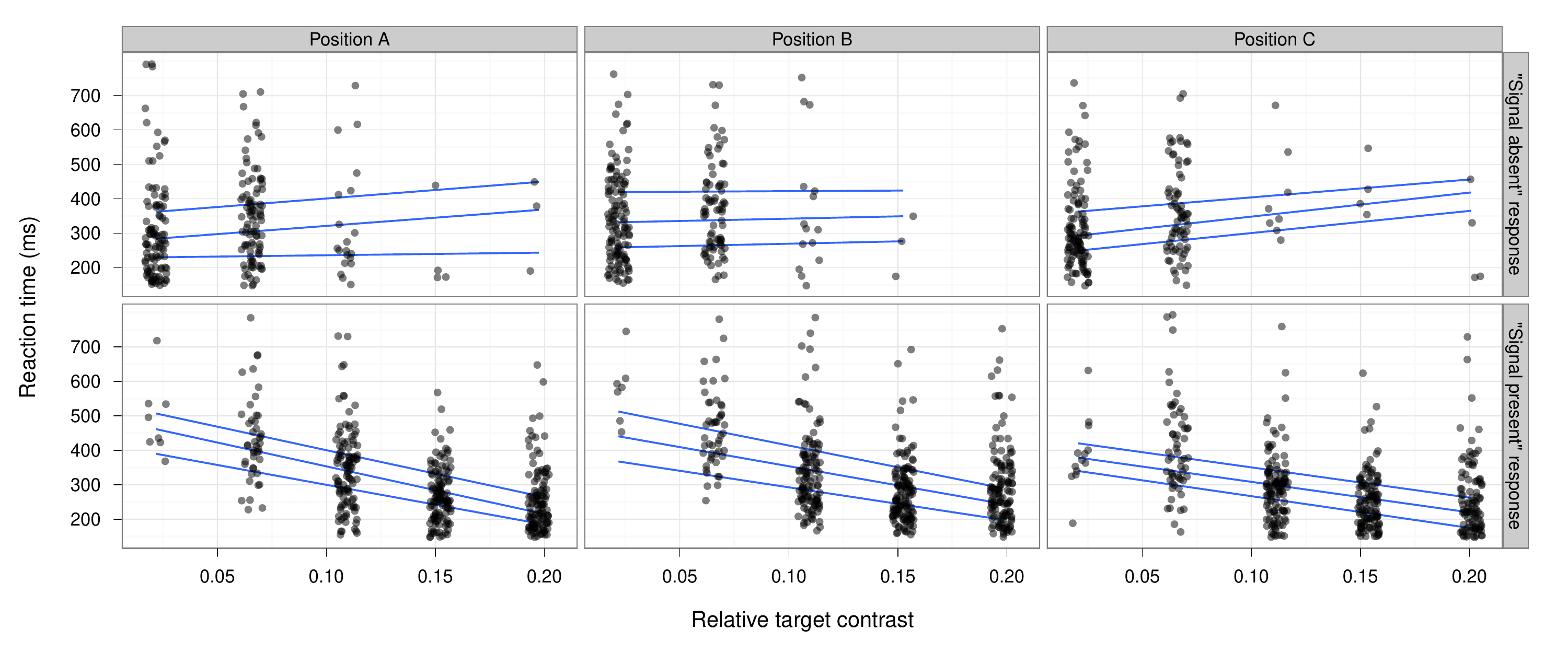}
\par\end{centering}

}

\caption{\selectlanguage{english}%
Choice (\textbf{a}) and reaction time (\textbf{b})\textbf{ }data in
a detection experiment by Maertens.\label{fig:RT-data} \selectlanguage{british}%
}
\end{figure}

We apply this model to data from an unpublished experiment by M. Maertens
(personal communication to Simon Barthelmé). The dataset is made of
1860 observations, obtained from a single human subject, which had
to choose between $k=2$ alternatives: ``signal absent'' (no light
increment was presented), or ``signal present'' (a light increment
was presented), under 15 different experimental conditions: 3 different
locations on the screen, and 5 different contrast values. Following
common practice in this field, trials with very high or very low reaction
times (top and bottom 5\%) were excluded from the dataset, because
they have a high chance of being outliers (fast guesses, keyboard
errors or inattention). The data are shown on Figure \ref{fig:RT-data}. 

From the description above, one sees that five parameters, $(m_{1},m_{2},c_{1},c_{2},s)$,
are required to describe the random behaviour of a single pair $(r_{i},d_{i})$,
when $k=2.$ To account for the heterogeneity introduced by the varying
experimental conditions, we assume that the 2 accumulation rates,
$m_{1}$, $m_{2}$ vary across the 15 experimental conditions, while
the 3 parameters related to the boundaries, $c_{1}$, $c_{2}$ and
$s$, are shared across conditions. The parameter $\bm{\theta}$ is
therefore 33-dimensional.

We note that this model would present a challenge for inference even
if the likelihood function was available. It is difficult to assign
priors to the parameters, because they do not have a clear physical
interpretation, and available prior knowledge (e.g., that reaction
times will normally be less than 1 second) does not map easily unto
them. Moreover, the model is subject in certain cases to weak identifiability
problems. For instance, if one response dominates the dataset, there
is little information available beyond the fact that one drift rate
is much higher than the other (or one threshold much lower than the
other, or both). 

We re-parametrised the positive parameters $c_{1}$, $c_{2}$ as $c_{1}=e^{\lambda}$,
$c_{2}=e^{\lambda+\delta}$, and assigned a $\mathcal{N}\left(0,1\right)$
prior to $\lambda$, $\delta$, and $s$. Taking a $N(0,5^{2})$ prior
for these 3 quantities led to similar results. Some experimentation
suggested that the drift rates could be constrained to lie between
-0.1 and 0.1, because values outside of this interval seem to yield
improbable reaction times (too short or too long). We assigned a $[-0.1,0.1]$
uniform prior for the 30 drift rates, and applied the appropriate
transform, i.e. $x\rightarrow\Phi^{-1}(0.1+5x)$, in order to obtain
a $N(0,1)$ prior for the transformed parameters. 

After a few unsuccessful attempts, we believe that this application
is out of reach of normal ABC techniques. The main difficulty is the
choice of the summary statistics. For instance, if one takes basic
summaries (e.g. quartiles) of the distribution of reactions times,
under each of the 15 experimental conditions, one ends with a very
large vector $\bm{s}$ of summary statistics. Due to the inherent
curse of dimensionality of standard ABC \citep{Blum:ABCNonParametricPerspective},
sampling enough datasets, of size 1860, which are sufficiently close
(in terms of the pseudo-distance $\left\Vert \bm{s}(\bm{y})-\bm{s}(\bm{y}^{\star})\right\Vert $)
would require enormous computational effort. Obviously, taking a much
smaller set of summary statistics would on the other hand lead to
too poor an approximation. 

Some adjustments are needed for ABC-EP to work on this problem. First,
notice that within a condition the datapoints are IID so that the
posterior distribution factorises over IID ``blocks''. We can therefore
employ the recycling technique described in Section \ref{sec:IID_case}
to save simulation time, by going through the likelihood sites block-by-block.
Second, since datapoints take values in $\{1,2\}\times\mathbb{R}^{+}$,
we adopt the following set of ABC constraints: $\I\left\{ d_{i}=d_{i}^{\star}\right\} \I\left\{ \left|\log r_{i}-\log r_{i}^{\star}\right|\le\epsilon\right\} $,
where $y_{i}^{\star}=(d_{i}^{\star},r_{i}^{\star})$ denotes as usual
the actual datapoints. Third, we apply the following two variance-reduction
techniques. One stems from the fact that each site likelihood does
not depend on the whole 33 parameters but on a subset of size 5. In
that case, using simple linear algebra, one can see that it is possible
to update only the marginal distribution of the EP approximation with
respect to these 5 parameters; see Appendix B for details. The second
is a simple Rao-Blackwellisation scheme that uses the fact that the
non-decisional component $r_{nd}$ is uniformly distributed, and may
therefore be marginalised out when computing the EP update. 

We report below the results obtained from ABC-EP with $\epsilon=0.16$,
$M=3\times10^{3}$, $\alpha=0.4$ (see end of Section \ref{sub:Numerical-stability}),
and 3 complete passes over the data; CPU time was about 40 minutes.
Results for smaller values of $\epsilon$, e.g. $\epsilon=0.1$, were
mostly similar, but required a larger CPU time.

Since we could not compare the results to those of a standard ABC
algorithm, we assess the results through posterior predictive checking\emph{.
}For each of the 15 experimental conditions,\emph{ }we generate 5,000
samples from the predictive density, and compare the simulated data
with the real, as follows. 

\begin{figure}
\begin{centering}
\includegraphics[scale=0.5]{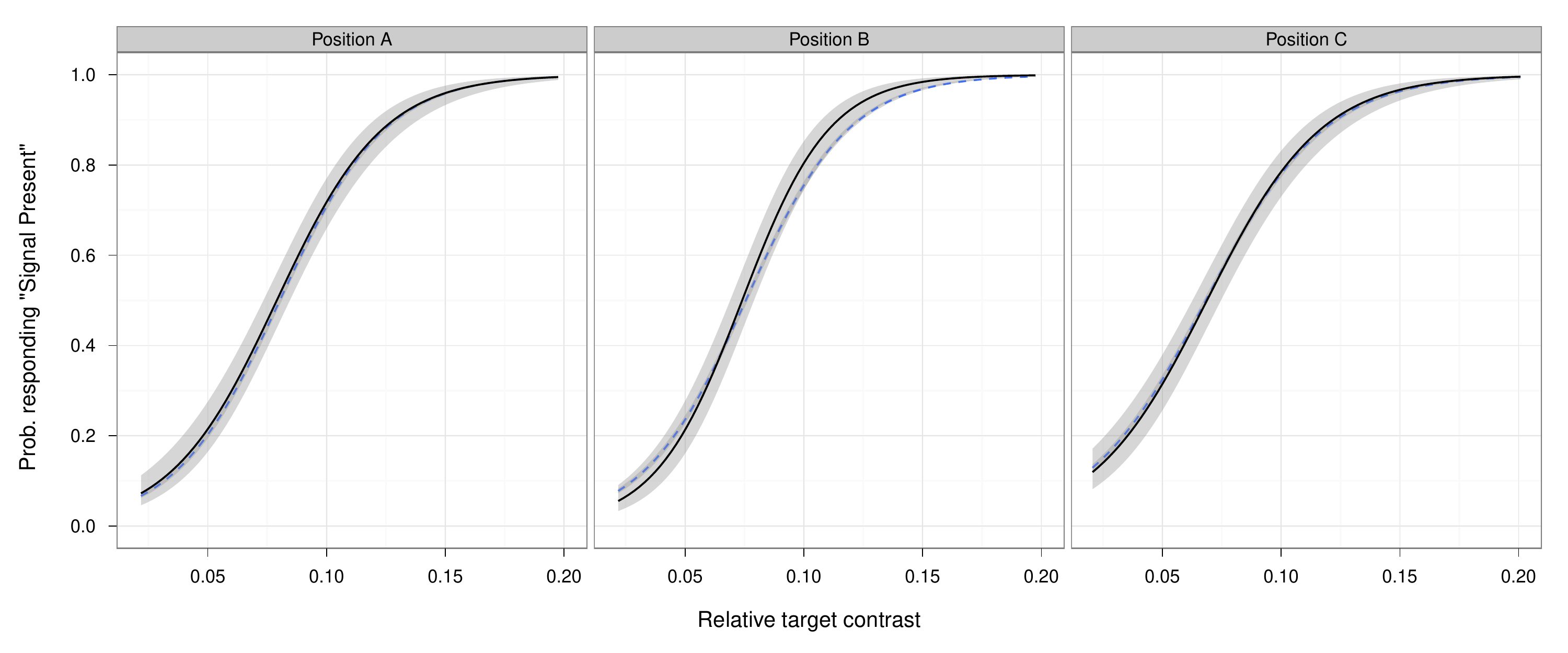}
\par\end{centering}

\caption{\selectlanguage{english}%
Probability of answering ``Signal Present'' as a function of contrast:
data versus (approximate) posterior predictive distribution. We sampled
the posterior predictive distribution and summarised it using a binomial
GLM with Cauchit link, in the same way as we summarised the data (see
(a) in Figure \ref{fig:RT-data}). The data are represented as a continuous
line, the predictive distribution as dotted. The posterior predictive
distribution is very close to the data. The shaded area corresponds
to a 95\% confidence interval for the fit to the data.\label{fig:Probability-sig-present-data-pred}\selectlanguage{british}%
}
\end{figure}

The marginal distribution of responses can be summarised by regressing
the probability of response on stimulus contrast, separately for each
stimulus position (as on Figure \ref{fig:RT-data}), and using a binomial
generalized linear model (with Cauchit link function). Figure \ref{fig:Probability-sig-present-data-pred}
compares data and simulations, and shows that the predictive distribution
successfully captures the marginal distribution of responses. 

\begin{figure}
\begin{centering}
\includegraphics[scale=0.5]{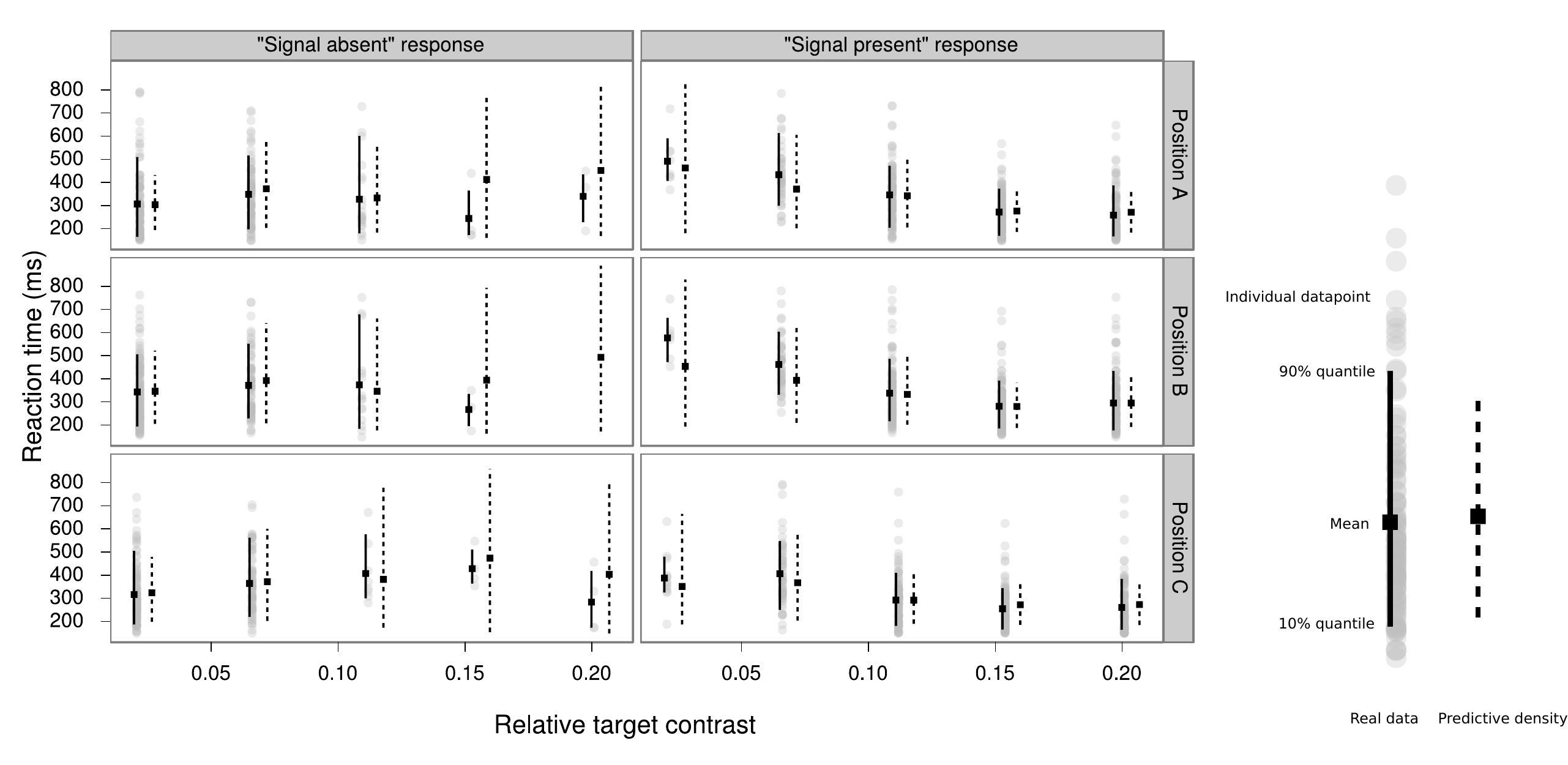}
\par\end{centering}

\caption{\selectlanguage{english}%
Reaction time distributions conditional on decision: data versus posterior
predictive distributions. The reaction times conditioned on contrast,
position and response are shown as grey dots and summarised via mean
and 10-90\% inter-quantile range (continuous lines). The posterior
predictive distributions computed from samples are summarised and
shown with an offset to the right (dotted lines). The conditional
densities are well captured, given sufficient data. \label{fig:Conditional-reaction-time-data-pred}\selectlanguage{british}%
}
\end{figure}

To characterise the distribution of reaction times, we look at means
and inter-quantile intervals, conditional on the response and the
experimental conditions. The results are presented on Figure \ref{fig:Conditional-reaction-time-data-pred}.
The predictive distributions capture the location and scale of the
reaction time distributions quite well, at least for those conditions
with enough data. Such results seem to indicate that, at the very
least, the ABC-EP approximate posterior corresponds to a high-probability
region of the true posterior.

\section{Difficult posteriors\label{sec:Difficult-posteriors}}

One obvious cause for concern is the behaviour of EP-ABC when confronted
with difficult posterior distributions, and in this section we explore
a few possible scenarios and suggest methods for detecting and correcting
potential problems. Of all potential problematic shapes a posterior
distribution could have, the worst is for it to have several modes,
and we deal with this question first. Nonmultimodal but otherwise
problematic posteriors are discussed later.

\subsection{Multimodality}

\begin{figure}
\begin{centering}
\includegraphics[scale=0.4]{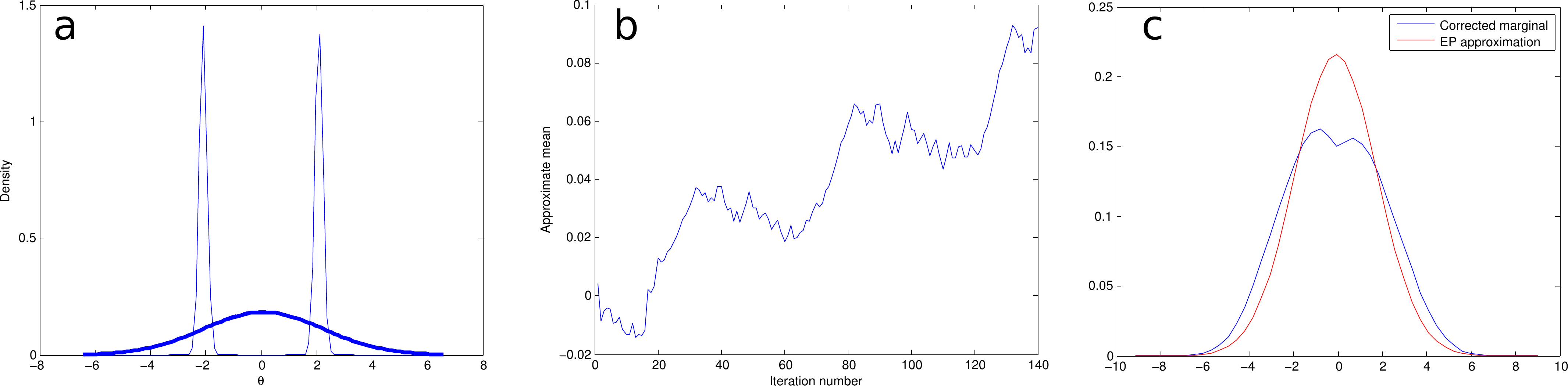}
\par\end{centering}

\caption{\label{fig:Multi}Multi-modal example. \textbf{a.} True posterior
pdf (thin line) versus EP-ABC approximation (thick line). \textbf{b.}
mean of the EP-ABC Gaussian approximation during the course of the
algorithm (i.e. vs iteration; 3 passes are performed, hence 150 iterations
in total) \textbf{c. }EP approximation vs. 1st-order PWO correction
(see Appendix }
\end{figure}

We begin with a toy example of a multimodal posterior. Consider the
following IID model: $y_{i}|\theta\sim N(|\theta|,1)$, $i=1,\ldots,n=50$,
and $\theta\sim N(0,10^{2})$. The dataset is obtained by sampling
from the model, with $\theta=2$. The true posterior may be computed
exactly, and is plotted in the left hand side of Figure \ref{fig:Multi}.
It features two symmetric, well separated modes, around $-2$ and
2. The Gaussian pdf $q$ that minimises $KL(q||\pi)$ where $\pi$
is the posterior (or, in other words, the Gaussian pdf, with mean
equal to posterior expectation, variance equal to the posterior variance)
is also represented, as a dashed line, but it cannot be distinguished
from the EP-ABC approximation (thick line). 

The behaviour of EP-ABC in this case is instructive. If we run the
standard EP-ABC algorithm (using the IID version, see Section 4 of
the paper, and with $M=8\times10^{6}$, $ESS_{\min}=500$), we obtain
a result very close to the the thick line in one pass (one cycle over
all the observations). However, if the algorithm is run for two passes
or more, there is a very high probability that the algorithm stops
its execution before completion, because many site updates will produce
negative definite contributions $\bm{Q}_{i}$. To perform more than
one pass, we have to use slow updates (as described in Section 3 of
the paper), and set $\alpha$ to a small value. The thick line in
Figure \ref{fig:Multi} was obtained with three passes, and slow updates
with $\alpha=0.1$. The right hand side of \ref{fig:Multi} shows
the type of plot one may use to assess convergence of EP-ABC: it represents
the evolution of the mean of the EP approximation along the iterations;
one ``iteration'' corresponds to one site update, and since we have
performed three passes over $n=50$ sites, there are 150 iterations
in total. This plot seems to indicate that the posterior expectation
is slowly drifting to the right, and does not stabilise. If we try
to run for more passes, we end up again with non-invertible covariance
matrices. This behaviour was already described by Minka in his PhD
thesis \citep{Minka:FamilyAlgApproxBayesInf} for the deterministic
case.

The first test to see if the posterior distribution is reasonable
is simply to run EP-ABC and see if it fails. There are two possible
causes for failures: either one used too few samples and Monte Carlo
variance led to negative definite updates, or the model itself is
problematic for EP, in which case EP will still fail when using large
sample sizes. 

Beyond this rather informal test there are more principled things
one could do. In \citet{Paquet:PerturbationCorrectionsApproxInference}
a very generic class of corrections to EP is introduced, and is described
in Appendix \ref{sec:PWOcorrections}. Their first-order correction
can be obtained relatively cheaply from the hybrid distributions,
and can be used to build corrected marginals. In Figure \ref{fig:Multi}
we show the first-order correction for our toy multi-modal example:
although the corrected marginal is still far from the true posterior,
it is clearly bimodal and in this case serves as a warning that the
Gaussian approximation is inappropriate (we also applied the same
type of correction in our third example, for which no MCMC gold standard
is available, but the correction did not modify in any noticeable
manner the marginal posterior distributions) In a similar vein, one
could use goodness-of-fit tests to see if the $n$ hybrid distributions
show large deviations from normality. 

Once the problem has been diagnosed, what can one do? The answer is,
unfortunately, ``not much''. If one can determine that multimodality
is caused by ambiguity in one or two parameters, it is possible to
effectively treat these parameters as hyperparameters. One separates
$\bt$ into $\bt_{A},\bt_{B}$ where $\bt_{b}$ is the set of problematic
parameters. Running EP-ABC with $p(\bt_{A}|\bm{y},\bt_{B})$ will
produce an estimate of $p(\bt_{b}|\bm{y})$ which can be used to form
an approximation to the marginals in a manner analoguous to the procedure
used in INLA \citep{rue2009approximate}. Although this will require
running EP-ABC several times for different values of $\bt_{B}$, it
might still be a lot cheaper than a MCMC procedure. 

If no information is available about what the cause of the multimodality
is or where the different modes are, our opinion is that all existing
ABC methods will have tremendous difficulties. Work might be better
invested in injecting more prior information or changing the parameterisation
such as to ensure there is only one mode.

\subsection{Difficult unimodal posteriors}

Compared to the toy example of the previous section, a more realistic
scenario is a posterior distribution that is unimodal but otherwise
still badly behaved. There are no theoretical results available on
convergence in this case, so we evaluated EP's behaviour in a case
of a rather nasty posterior distribution, adapted from the model of
reaction times described above.

We devised a problem in which we guessed the posterior would be at
the minimum very badly scaled and would probably have an inconvenient
shape. In the context of the reaction times model described in Section
\ref{sub:Reaction-times}, imagine that in a choice experiment the
subject picked exclusively the second category; so that we have no
observed reaction times for the first category. From the point of
the model this occurs whenever the threshold for the first category
is high enough, compared to accumulation speed, that the second accumulator
always wins. This creates complete uncertainty as to the ratio of
accumulation speed vs. threshold value for the first accumulator. 

\begin{figure}
\begin{centering}
\includegraphics[scale=0.5]{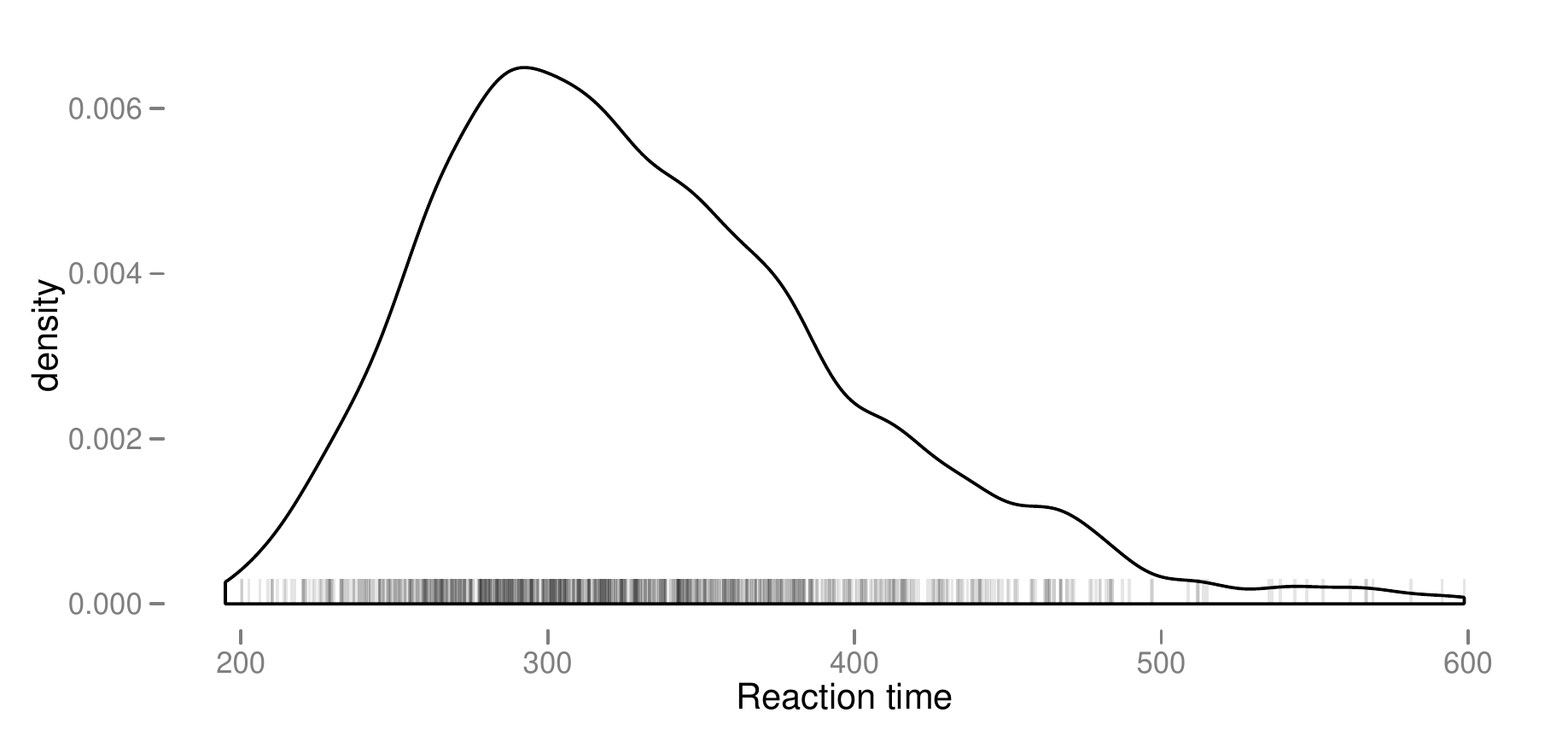}
\par\end{centering}

\caption{Simulated reaction time data used in creating the difficult posterior
of Section \ref{sec:Difficult-posteriors}. Individual reaction times
are marked with ticks, and a kernel density estimate is overlaid.
\label{fig:Difficult-posterior-data}}
\end{figure}

We generated a large dataset (1,000 datapoints) based on fixed values
of $(m_{1},m_{2},c_{1},c_{2},s)$: $m_{1}=10^{-3},\, m_{2}=0.08,\, c_{1}=20,\, c_{2}=10,$
and $s=0$. In this dataset there are no decisions favouring the first
category, and the resulting reaction time distribution is plotted
on Fig. \ref{fig:Difficult-posterior-data}. To make the inference
problem still manageable using MCMC-ABC, we limit inference to three
parameters: $\log m_{1}$, $\log m_{2}$ (log-accumulation rate of
the two accumulators) and $\log c_{1}$ (log-threshold for the first
category). The other two parameters are considered known and fixed
at their true value. We chose these parameters because we expected
the posterior to show quasi-certainty for $\log m_{2}$ and very high
uncertainty for $\mbox{log}\frac{m1}{c_{1}}$. To further increase
uncertainty, we chose a very vague prior $\bt\sim\N\left(0,10^{2}\times\bm{I}\right)$. 

We ran EP-ABC on this problem using the recycling strategy described
in section \ref{sub:Numerical-stability} (data are IID). In this
case EP-ABC needs large sample sizes to be stable numerically, which
is probably due to the very ill-conditioned covariance matrices that
arise (itself due to a very poorly-scaled posterior, more on that
below). To eliminate any such problems we deliberately chose to use
a very high number of samples, 3 million, with a minimum expected
sample size of 500. We used an acceptance window of 10ms. We did 3
passes over the data, for a total computing time of 9 minutes on a
standard laptop. To check for stability, we ran EP-ABC 10 times. 

To check the accuracy of the results, we ran a MCMC-ABC sampler that
used as its summary statistics 8 quantiles of the reaction time distribution,
linearly spaced from 0.01 to 0.99, and whether the first category
was ever chosen. Quantiles were rescaled by a factor of $\frac{1}{200}$.
Samples were accepted if the second category was chosen less than
100\% of the time, or if the Euclidean distance between $\left|\bm{s}\left(\bys\right)-\bm{s}\left(\by\right)\right|$
was over $\epsilon=0.025$. This latter value was found by iteratively
adjusting to get an acceptance rate of 1/1000. 

\begin{center}
\begin{figure}
\begin{centering}
\includegraphics[scale=0.35]{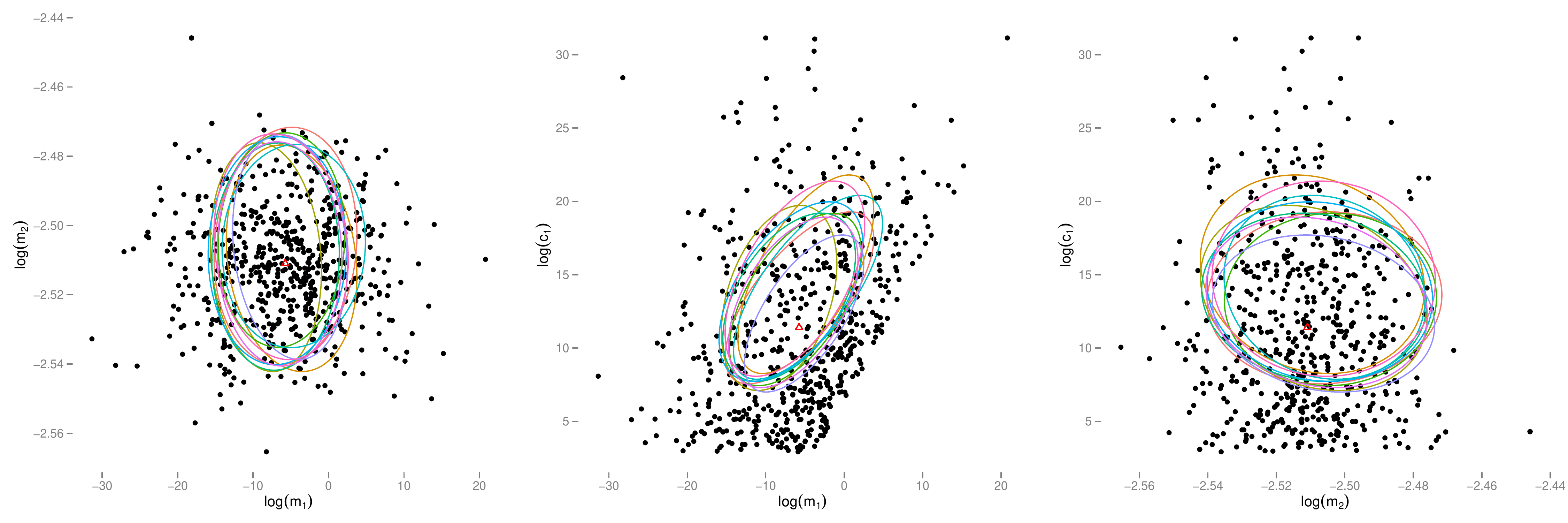}
\par\end{centering}

\caption{Two-dimensional marginal distributions of the posterior distribution
for section \ref{sec:Difficult-posteriors}. In black, MCMC-ABC samples.
The colored ellipses represent contours for EP-ABC approximate posteriors
(scaled to span 4 times the standard deviation across the diameter).
Each ellipse represents the results for a different run of EP-ABC.
The posterior distribution exhibits extremely poor scaling (compare
the range of $\log m_{1}$ to that of $\log m_{21}$), and a truncated
shape along some dimensions. Red triangles represent posterior mean
computed from MCMC-ABC samples. \label{fig:Difficult-posterior-results}}
\end{figure}

\par\end{center}

We would like to stress that an appropriate proposal distribution
was found using the covariance matrix obtained using ABC-EP (and readjusted
later based on short runs). The scaling of the posterior is such that
using, e.g., the prior covariance matrix in a proposal distribution
is impossible: one ends up using extremely high values of $\epsilon$,
and therefore with a very poor approximation of the posterior. As
a starting value we used the true parameters. After several adjustments,
we ran a MCMC-ABC chain for 3 million iterations, which we trimmed
down to 600 samples for visualisation on Fig. \ref{fig:Difficult-posterior-results}
(auto-correlation in the chain being in any case extremely high).

MCMC-ABC samples are shown on Fig. \ref{fig:Difficult-posterior-results},
along with density contours representing EP-ABC results. Although
MCMC-ABC and EP-ABC target different densities, it is reasonable to
hope that in this case these densities must be fairly close, since
we deliberately set the acceptance ratio to be very small, and the
data are simple enough for the summary statistics to be adequate.
The posterior distribution obtained using MCMC-ABC, as shown on figure
\ref{fig:Difficult-posterior-results} is pathological, as expected.
Scaling is very poor: one parameter varies over a very small range
$(-2.56,-2.44)$ while the other two vary over $[-30,20]$ and $[0,30]$.
The posterior also has a difficult shape and appears truncated along
certain dimensions. EP-ABC provides approximations that are located
squarely within the posterior distribution, but seems to underestimate
posterior variance for the third parameter, $\log c_{1}$. Given the
difficult context, EP-ABC still performs arguably rather well. At
the very least it may be used to find an appropriate proposal distribution
for MCMC-ABC, which may perform extremely poorly without such help.

\section{Extensions\label{sec:Extensions}}

Our description of EP-ABC assumes the prior is Gaussian, and produces
a Gaussian approximation. With respect to the latter point, we have
already mentioned (see also Appendix A) that EP, and therefore EP-ABC,
may propagate more generally approximations from within a given exponential
family; say the parametric family of Wishart distributions if the
parameter is a covariance matrix. Regarding the prior, even when considering
Gaussian approximation, it is possible to accommodate a non Gaussian
prior, by treating the prior as an additional site. 

Thus, the main constraint regarding the application of EP-ABC is that
the likelihood is factorizable, i.e. $p(\by|\bt)=\prod_{i=1}^{n}p(y_{i}|y_{1:i-1},\bt)$
in such a way that simulating from each factor is feasible. 

In this paper, we focused our attention on important applications
of likelihood-free inference where the likelihood is trivial to factorise;
either because the datapoints are independent, or Markov. But ABC-EP
is not limited to these two situations. First, quite a few time series
models may be simulated sequentially, even if they are not Markov.
For instance, one may apply straightforwardly EP-ABC to a GARCH-stable
model \citep[e.g.][]{liu1995maximum,mittnik2000diagnosing,menn2005garch},
which is a GARCH model \citep{bollerslev1986gac} with an alpha stable-distributed
innovation. Second, one may obtain a simple factorisation of the likelihood
by incorporating latent variables into the vector $\bm{\theta}$ of
unknown parameters. Third, one may replace the true likelihood by
a certain approximation which is easy to factorise. The following
section discusses and illustrates such an approach, based on the concept
of composite likelihood.

\subsection{Composite likelihood}

Composite likelihood is an umbrella term for a family of techniques
based on the idea of replacing an intractable likelihood by a factorisable
pseudo-likelihood; see the excellent review of \citet{varin2011overview}.
For the sake of simplicity, we focus on marginal composite likelihood,
but we note that other versions of composite likelihood (where for
instance the factors are conditional distributions) could also be
treated similarly. 

In our EP-ABC context, replacing the likelihood by a marginal composite
likelihood leads to the following type of ABC posterior
\[
p_{\varepsilon}^{CL}(\bt|\bm{y})\propto p(\bt)\prod_{s=1}^{n_{s}}\int p(\by_{s}|\bt)^{\eta_{s}}\I_{\left\{ \Vert\by_{s}-\bm{y}_{s}^{\star}\Vert\leq\varepsilon\right\} }\, d\by_{s}
\]
where $p(\by_{s}|\bt)$ is the marginal distribution of some subset
$\bm{y}_{s}$ of the observations, and $\eta_{s}$ is a non-negative
weight. There exists some theory on how to choose the weights $\eta_{s}$
so that the composite likelihood is close enough to the true likelihood
in some sense, see again \citet{varin2011overview}, but for simplicity
we shall take $\eta_{s}=1$. Clearly, EP-ABC may be applied straightforwardly
to this new ABC posterior, provided one may sample independently from
the $n_{s}$ marginal distributions $p(\by_{s}|\bt)$. In fact, the
$n_{s}$ factors may be treated as IID factors, which makes it possible
to use the recycling strategy described in Section \ref{sec:IID_case}.

To make this idea more concrete, we consider the class of hidden Markov
models, where the datapoints $y_{i}$ are conditionally independent,
$y_{i}|x_{i},\bt\sim g_{\bt}(y_{i}|x_{i})$, conditional on latent
variable $x_{i}$, and the $x_{i}$'s are Markov: $x_{1}\sim\mu_{\bt}(x_{1})$,
$x_{i+1}|x_{1:i},\bt\sim f_{\bt}(x_{i+1}|x_{i})$ for $i\geq1$; see
\citet{andrieu2005line} for a previous application of composite likelihood
to hidden Markov models. We assume assume that the density $g_{\bt}(y_{i}|x_{i})$
is intractable, and that one may sample from it; see e.g. \citet{dean2011parameter},
\citet{calvet2011state} for applications of likelihood free inference
to this class of intractable hidden Markov models. We assume that
$\mu_{\bt}$ is the stationary distribution of the Markov process;
hence marginally $x_{i}|\bt\sim\mu_{\bt}(x_{i})$. 

Then a particular version of the composite likelihood ABC posterior
may be constructed as follows: for some fixed integer $l\geq2$, take
$n_{s}=\left\lceil n/l\right\rceil $ and $\by_{s}=y_{l(s-1)+1:ls}$
if $ls\leq n$, $\by_{s}=y_{l(s-1)+1:n}$ otherwise. One may sample
from $p(\by_{s}|\bt)$ as follows: sample $x_{l(s-1)+1}|\bt\sim\mu_{\theta}(x_{l(s-1)+1})$,
then sample recursively $x_{i+1}|x_{i},\bt\sim f_{\bt}(x_{i+1}|x_{i})$
for $i=l(s-1)+2,\ldots,ls$, and finally sample $y_{i}|x_{i},\bt\sim g_{\bt}(y_{i}|x_{i})$
independently for $i=l(s-1)+1,\ldots,ls$. 

As an illustration, we consider the following alpha-stable stochastic
volatility model: $x_{1}\sim N\left(\mu,\sigma^{2}/(1-\rho^{2})\right)$,
$x_{i+1}=\mu+\rho(x_{i}-\mu)+\sigma u_{t}$, with $u_{t}\sim N(0,1)$,
and $y_{i}|x_{i},\bt$ is a Stable distribution with the following
parameters (using the same parametrisation as in Section \ref{sub:First-example:-Alpha-stable}):
$\alpha\in(1,2)$ is fixed, $\beta=0$ (no skewness), $\delta=0$
(centred at zero),  and the scale parameter $\gamma$ is set to $\exp(x_{t}/2)$.
The parameter vector is therefore $\bt=\left(\mu,\Phi^{-1}((\rho+1)/2),\log\sigma,\Phi^{-1}(\alpha-1)\right)$,
and we use the following prior, $\bt\sim N\left((0,1.65,0,0)^{T},\mathrm{diag}\left(100,0.25,1,1\right)\right)$;
for the second and third components (corresponding to $\rho$ and
$\sigma$), we fitted a Gaussian distribution to the prior suggested
by \citet{kim1998stochastic} (after the appropriate transformation),
while for the fourth component, the prior is equivalent to $\alpha\sim U[1,2]$. 

\begin{figure}[h]
\begin{centering}
\includegraphics[scale=0.5]{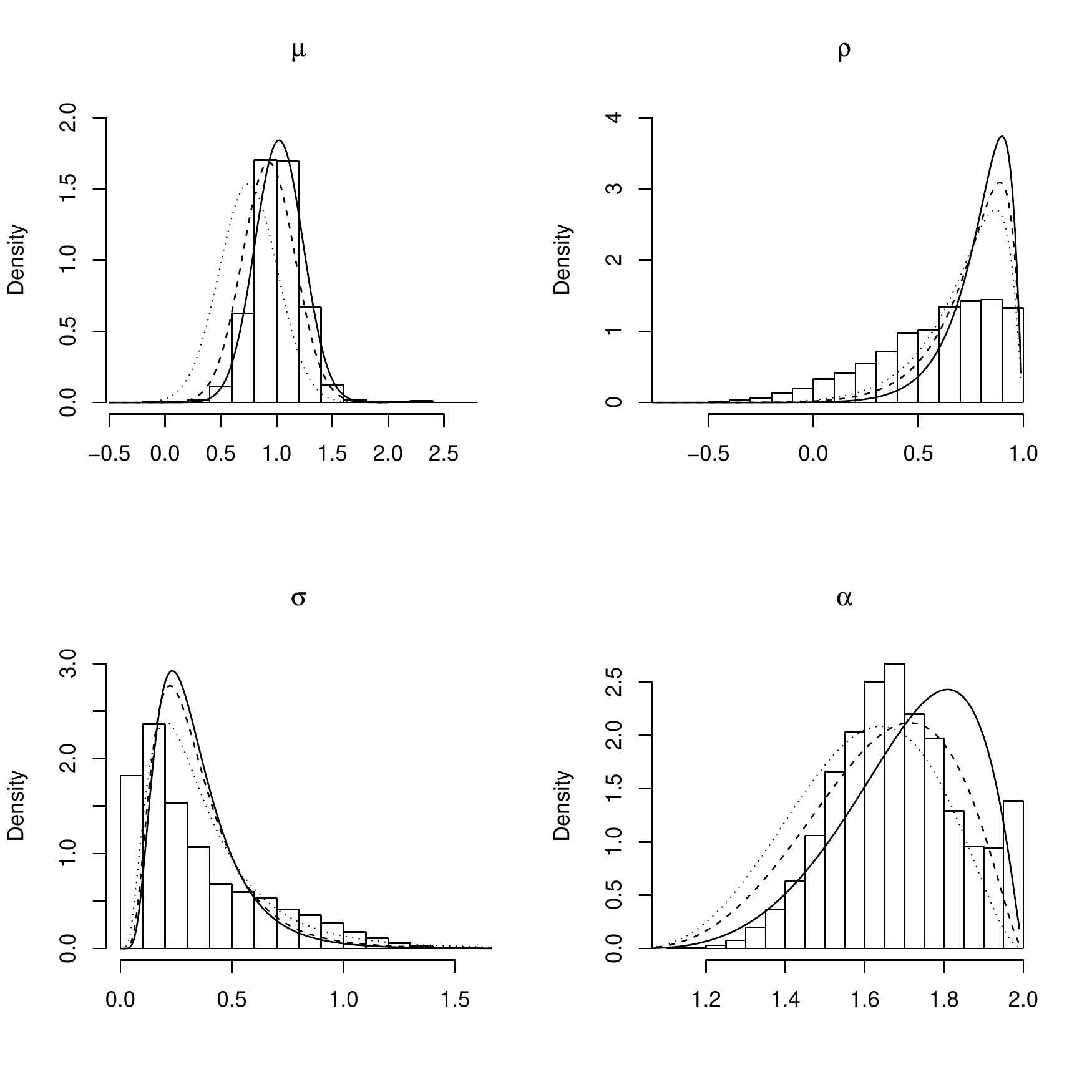}
\par\end{centering}

\caption{\label{fig:HMM-SV}Comparison of EP-ABC approximations based on the
composite likelihood for block sizes $L=2$ ($\varepsilon=1$, solid
line), $L=3$ ($\varepsilon=1.5$, dashed line), $L=4$ ($\varepsilon=2.5$,
dotted line) and output (histograms) from a PMCMC sampler targeting
an ABC posterior corresponding to $n$ constraints $|y_{i}-y_{i}^{\star}|<\varepsilon$,
and $\varepsilon=1$. }
\end{figure}

We simulated a dataset of size $n=120$ from this model, with true
parameters $\mu=0.35$, $\sigma=0.2$, $\rho=0.97$, $\alpha=1.5$.
Note that the high value of $\rho$ creates strong correlations between
successive blocks, so it is interesting to see if the marginal composite
likelihood remains a reasonable approximation of the true likelihood
in this case. We considered several values of $l$: $l=2,3,4$. The
choice of $l$ amounts to a trade-off between the accuracy of the
composite likelihood approximation (the larger $l$, the better),
and the computational cost (the larger $l$, the smaller the probability
of the event $ $$\Vert\by_{s}-\bm{y}_{s}^{\star}\Vert\leq\varepsilon$).
In practice, we observe that it is difficult to take $l$ to be very
large, because the probability that $\Vert\by_{s}-\bm{y}_{s}^{\star}\Vert\leq\varepsilon$
decrease exponentially in $l$. For each value of $l$, we took $\epsilon$
to be as small as possible, subject to the running time being about
one minute and a half (and roughly about $10^{7}$ draws from an alpha-stable
distribution); thus, $\varepsilon=1$, $1.5,$ $2.5$ for $l=2$,
$3,$ $4$. Fig. \ref{fig:HMM-SV} plots the marginal distributions
for each component, obtained by averaging out 10 runs of ABC-EP and
applying the appropriate transformation. These marginal densities
are to be compared with the output of a PMCMC algorithm that targets
the ABC posterior that correspond to the $n$ constraints $\left|y_{i}-y_{i}^{\star}\right|\leq\varepsilon$,
with $\varepsilon=1$. (This PMCMC algorithm is a random walk Metropolis
sampler in the $\bt$-dimension, which runs the ABC filtering algorithm
of \citet{filteringabc} at each iteration.) The running time of that
PMCMC algorithm was three days and six hours ($10^{5}$ iterations,
$N=5\times10^{3}$ particles, leading to $6\times10^{10}$ draws from
an alpha-stable distribution). Relative to previous examples, the
approximation error brought by EP-ABC-CL is more noticeable in this
case (especially for $\rho$), but remains quite reasonable. Presumably
the main source of error is the composite likelihood approximation.
Note also that the PMCMC output should not considered as a gold standard
in this case, as it corresponds to an ABC approximation based on a
different set of constraints. 

The complexity of EP-ABC-CL in this case is $O(n)$, while the complexity
of PMCMC is $O(n^{2})$ \citep{PMCMC}. Thus it is easy to apply EP-ABC-CL
to datasets of size $n=10^{3}$, or even $10^{4}$, while this would
prove very expensive for PMCMC. 

Apart from hidden Markov models, there are several other classes of
models that could be treated using the composite likelihood version
of EP-ABC. For instance, it is common to use marginal composite likelihood
to deal with spatial extremes \citep[see e.g.][]{SpatialExtremesReview};
however only bivariate marginal distributions are tractable in such
a case, whereas with EP-ABC, one could deal with composite likelihood
made of larger-order marginals. 

\section{Conclusion\label{sec:Conclusion}}

EP-ABC has several limitations. It requires the likelihood of the
model to be factorisable. It produces a parametric approximation of
the posterior, which may be a poor approximation in certain cases;
e.g. a Gaussian approximation while the posterior is severely multimodal
(although one may wonder which ABC method would work under such a
scenario.) And the mathematical properties of EP-ABC (i.e. convergence,
assessment of the approximation error) are not yet fully understood.
Work on EP has started recently \citep{mengersen2011mixtures}, but
these preliminary results do not address the issue of the stability
of the algorithm when each site update introduces a stochastic error,
as in EP-ABC. This is certainly an important (but possibly quite difficult)
direction for future research. 

As of now, we would like to make two pragmatic remarks. First, EP-ABC
is very fast, compared to other ABC methods. We have observed empirically
that it produces very accurate results, but the user is free to either
use these results directly, or a first step in order to calibrate
a second, more expensive step based on a standard ABC approximation.
This type of calibration is often critical to obtain decent convergence
in standard ABC. Second, EP-ABC greatly reduces the pain of designing
summary statistics (as only site-specific summary statistics $s_{i}$
must be chosen in EP-ABC), and in certain cases make it possible to
do away completely with summary statistics (i.e. take $s_{i}(y_{i})=y_{i}$
at each site). This seems quite convenient as, in real-world applications,
 one has little intuition and even less mathematical guidance on to
why $p(\bt|\bm{s}(\by))$ should be close to $p(\bt|\by)$ for a given
set of summary statistics $\bm{s}$. One may even argue that the dependence
on summary statistics is currently the main limitation of the ABC
approach, and that it is essential that this issue is addressed in
future research in likelihood-free inference, whether through EP or
by other means.

\section*{Acknowledgements}

The authors thank the associate editor, the referees, Pierre Jacob,
Jean-Michel Marin, Pierre Pudlo, Christian P. Robert, Sumeetpal S.
Singh, Scott Sisson and Darren Wilkinson for helpful comments, and
M. Maertens for providing the data used in the third example. The
first author acknowledges support from the BMBF (Foerderkennzeichen
01GQ1001B). The second author acknowledges support from the ``BigMC''
ANR grant ANR-008-BLAN-0218 of the French Ministry of Research.

\section*{Appendix A: Algorithmic description of EP}

The more general EP algorithm for a generic exponential family is
described as Algorithm \ref{alg:Generic-EP}. The sites are therefore
of the form $f_{i}(\bt)=\exp\left\{ \bm{\lambda_{i}}^{t}\bm{t}\left(\bt\right)-\phi\left(\bm{\lambda}_{i}\right)\right\} $,
so that the global approximation pertains to the same family, i.e.
$q(\bt)=\exp\left\{ \bm{\lambda}^{t}\bm{t}(\bt)-\phi(\bm{\lambda})\right\} $,
$\bm{\lambda}=\sum_{i=0}^{n}\bm{\lambda}_{i}$. For a given exponential
family, there exists a smooth invertible mapping $\bm{\lambda}=\bm{\lambda}(\bm{\eta})$
that transforms the moment parameters $\bm{\eta}=\mathbb{E}_{\bm{\lambda}}\left\{ \bm{t}(\bt)\right\} =\int\bm{t}(\bt)\exp\left\{ \bm{\lambda_{i}}^{t}\bm{t}\left(\bt\right)-\phi\left(\bm{\lambda}_{i}\right)\right\} \, d\bt$,
into the natural parameters $\bm{\eta}$ \citep[see e.g. ][p. 105]{Schervish}.
In the particular case where the exponential family is the family
of Gaussian distributions of dimension $d$, as described in the paper,
one simply takes $\bl=(\bm{r},\bm{Q})$, and $\bm{\eta}=(\bm{\mu},\bm{\Sigma})$.
In particular, Step 2 of Algorithm \ref{alg:Generic-EP} corresponds
exactly to computing the moments in \eqref{eq:EP_moments_Gaussiancase}.

\begin{algorithm}
Input: a target density $\pi\left(\bt\right)=p\left(\bt\right)\prod_{i=1}^{n}l_{i}\left(\bt\right)$. 

Initialise $\bl_{0}$ to the exponential parameters of the prior $p_{0}$,
and local site parameters $\bl_{1}\ldots\bl_{n}=0$. Set global approximation
parameter $\bl$ to $\bl=\sum_{i=0}^{n}\bl_{i}=\bl_{0}$. Loop over
sites $i=1,\ldots,n$ until convergence:
\begin{enumerate}
\item Create hybrid distribution $h(\bt)\propto q_{-i}\left(\bt\right)l_{i}(\bt)$
by setting $q_{-i}\left(\bt\right)\propto\exp\left(\bl_{-i}^{t}\bm{t}\left(\bt\right)\right)$
with $\bl_{-i}=\bl-\bl_{i}$. 
\item Compute moments $\be_{h}$ of hybrid distribution, transform to natural
parameters $\bl_{h}=\bl(\bm{\eta}_{h})$. 
\item Update site $i$ by setting $\bl_{i}=\bl_{h}-\bl_{-i}$, then reset
global parameter $\bl$ to $\bl=\bl_{h}$. 
\end{enumerate}
Return moment parameters $\be=\be\left(\bl\right)$. 

\caption{\selectlanguage{english}%
Generic EP for exponential families.\label{alg:Generic-EP} \selectlanguage{british}%
}
\end{algorithm}

\section*{Appendix B: marginal EP updates\label{sub:Special-EP-updates}}

In example 2 we make use of the fact that some sites only depend on
a subset of the parameters to obtain more stable updates. We list
below some results for multivariate Gaussian families that are essential
in deriving these special EP updates. 

We generalise the problem slightly to computing the moments of a hybrid
$h\left(\bt\right)\propto f\left(\bm{A}\bt\right)\mathcal{N}\left(\bt;\bmu_{0},\bS_{0}\right)$,
the product of a multivariate Gaussian density and a likelihood which
is a function of $\bm{A}\bt$, where \textbf{$\bm{A}$ }is a matrix
of dimension $k\times m$, $m<k$. When $\bm{A}$ is a sub-matrix
of the identity matrix we have the special case of a likelihood which
only depends on a subset of the parameters. 

For the normalisation constant $Z$, we have that:

\[
Z=\int f\left(\bm{A}\bt\right)\mathcal{N}\left(\bt;\bmu_{0},\bS_{0}\right)\mbox{d}\bt=\int f\left(\bm{z}\right)\mathcal{N}\left(\bm{z},\bm{A}\bmu_{0},\bm{A}\bS_{0}\bm{A}^{\bm{t}}\right)\mbox{d}\bm{z}
\]
where $\bm{z}=\bm{A}\bt$. For the expectation of the hybrid:

\[
\frac{1}{Z}\int\bt f\left(\bm{A}\bt\right)\mathcal{N}\left(\bt;\bmu_{0},\bS_{0}\right)\mbox{d}\bt=\frac{1}{Z}\int f(\bm{z})\mathcal{N}\left(\bm{z},\bm{A}\bmu_{0},\bm{A}\bS_{0}\bm{A}^{\bm{t}}\right)E(\bt|\bm{z})\mbox{d}\bm{z}
\]
with $E(\bt|\bm{z})=\bm{Vz}+\bm{b}$, $\bm{V}=\bS_{0}\bm{A^{t}}\left(\bm{A}\bS_{0}\bm{A^{t}}\right)^{-1}$
and $\bm{b}=\bmu_{0}-\bS_{0}\bm{A^{t}}\left(\bm{A}\bS_{0}\bm{A^{t}}\right)^{-1}\bm{A}\bmu_{0}$,
thus
\[
\mathbb{E}_{h}(\bt)=\bm{V}\mathbb{E}_{h}(\bm{z})+\bm{b}
\]
where $\mathbb{E}_{h}(\bm{z})$ is the expectation of the hybrid.
A similar calculation yields an expression for the covariance:

\begin{equation}
Cov_{h}\left(\bt\right)=\bS_{0}-\bS_{0}\bm{A^{t}}\left(\bm{A}\bS_{0}\bm{A^{t}}\right)^{-1}\bm{A}\bS_{0}+\bm{V}Cov_{h}\left(\bm{z}\right)\bm{V^{t}}=\left(\bm{I}-\bm{VA}\right)\bS_{0}+\bm{V}Cov_{h}\left(\bm{z}\right)\bm{V^{t}}\label{eq:special_update_cov}
\end{equation}

These three results yield computational savings and increased stability,
because the moments of the hybrid distribution over $\bt$ can be
obtained from the moments of the marginal hybrid over $\bm{z}$, which
has lower dimensionality.

\section*{Appendix C: Paquet-Winter-Opper corrections\label{sec:PWOcorrections}}

In \citet{Paquet:PerturbationCorrectionsApproxInference} a method
for correcting an EP approximation is presented (denoted henceforth
the PWO correction). We give below a simpler way of deriving the corrections,
and show how to apply them in an ABC context.

Using the factorisations given in \eqref{eq:EPdecomposition} and
\eqref{eq:q-factorised}, the PWO correction may derived as 

\begin{equation}
\pi\left(\bt\right)=q\left(\bt\right)\prod_{i=1}^{n}\frac{l_{i}\left(\bt\right)}{f_{i}\left(\bt\right)}\stackrel{\Delta}{=}q\left(\bt\right)\prod_{i=1}^{n}\left\{ 1+e_{i}(\bt)\right\} \label{eq:target-reexpressed}
\end{equation}
where the $e_{i}$'s are error terms. This product can be expanded
in increasing orders of $e_{i}$: 

\[
\pi\left(\bt\right)=q(\bt)\left\{ 1+\sum_{i=1}^{n}e_{i}\left(\bt\right)+\sum_{j<k}e_{j}\left(\bt\right)e_{k}\left(\bt\right)+\ldots\right\} 
\]

The PWO corrections are obtained by truncating to a given order. Note
that truncation might result in a function that may not be everywhere
positive, although in practice the problem did not arise either in
the original applications \citep{Paquet:PerturbationCorrectionsApproxInference},
or in ours. 

The first-order (un-normalised) correction has a particularly simple
form:

\begin{eqnarray*}
q_{c}^{1}\left(\bt\right) & \propto & q(\bt)\left\{ 1+\sum_{i=1}^{n}e_{i}\left(\bt\right)\right\} \\
 & = & q(\bt)+\sum_{i=1}^{n}q(\bt)\left(\frac{l_{i}\left(\bt\right)}{f_{i}\left(\bt\right)}-1\right)\\
 & = & (n-1)q\left(\bt\right)+\sum_{i=1}^{n}q_{-i}(\bt)l_{i}\left(\bt\right)
\end{eqnarray*}
where in the second part of the expression we recognise the \emph{hybrid
}distributions $h_{i}\propto q_{-i}(\bt)l_{i}\left(\bt\right)$. The
integration constant of $q_{c}^{1}\left(\right)$ can easily be obtained
as a by-product of quantities computed during the EP run:

\[
\int q_{c}^{1}(\bt)\, d\bt=\left(n-1\right)Z_{q}+\sum_{i=1}^{n}Z_{i}
\]
where the $Z_{i}$ are the integration constants of the hybrids. 

As \citet{Paquet:PerturbationCorrectionsApproxInference} note, the
mean and covariance of the first-order approximation $q_{1}$ are
the same as that of $q$, provided that EP has converged, which by
definition happens when the hybrids of the $n$ sites have the same
expectation and covariance matrix, and are equal to respectively the
expectation and covariance matrix of the global approximation $q$.

We can still use the first-order correction for other expectations
$\mathbb{E}_{q_{c}^{1}}\left[f(\bt)\right]$: for example, using $f_{\alpha}(\bt)=\mathbb{I}\left(\theta_{j}<\alpha\right)$
for different values of $\alpha$ leads to an improved estimate of
the posterior marginal of $\theta_{j}$. (This is the strategy we
use to obtain the right panel in Figure \eqref{fig:Multi}.) The first-order
correction comes nearly for free in ABC-EP, but higher-orders seem
much more expensive to obtain in an ABC context, and are not discussed
here. 

\bibliographystyle{apalike}
\bibliography{bib}

\end{document}